\documentclass[final,5p,times,twocolumn]{elsarticle} 

\biboptions{sort&compress}

\graphicspath{{./figures/}}

\newcommand*{\bibliopath}{./biblio}

\newcommand*{\diagrampath}{./tikz}

\newcommand{\topology}{\ensuremath{T}}
\newcommand{\form}{\ensuremath{F}}
\newcommand{\force}{\ensuremath{F}^{*}}
\newcommand{\constrainedform}{\ensuremath{\bar{F}}}

\newcommand{\trails}{\ensuremath{\Omega}}
\newcommand{\trail}{\ensuremath{\omega}}

\newcommand{\auxiliarytrail}{\ensuremath{\omega^{\text{a}}}}

\newcommand{\sequence}{\ensuremath{k}}
\newcommand{\sequences}{\ensuremath{\bm{k}}}

\newcommand{\nodes}{\ensuremath{\mathcal{V}}}
\newcommand{\node}{\ensuremath{\mathbf{v}}}
\newcommand{\nodeorigin}{\ensuremath{\mathbf{v}^{\text{o}}}}
\newcommand{\nodesupport}{\ensuremath{\mathbf{v}^{\text{s}}}}

\newcommand{\supports}{\ensuremath{\mathcal{S}}}

\newcommand{\edges}{\ensuremath{\mathcal{E}}}
\newcommand{\edge}{\ensuremath{\mathbf{e}}}

\newcommand{\trailedge}{\ensuremath{\mathbf{e}^{\text{t}}}}

\newcommand{\deviationedge}{\ensuremath{\mathbf{e}^{\text{d}}}}

\newcommand{\internalforcestates}{\ensuremath{\mathbf{C}}}
\newcommand{\internalforcestate}{\ensuremath{c}}

\newcommand{\designparameters}{\ensuremath{\mathbf{x}}}

\newcommand{\equilibriumattributes}{\ensuremath{\mathbf{u}}}
\newcommand{\constrainedequilibriumattributes}{\ensuremath{\bar{\equilibriumattributes}}}

\newcommand{\energy}{\ensuremath{\eta}}
\newcommand{\energythreshold}{\ensuremath{\eta^{\text{min}}}}

\newcommand{\iteration}{\ensuremath{\tau}}
\newcommand{\iterationsmax}{\ensuremath{\tau^{\text{max}}}}

\newcommand{\edgeforces}{\ensuremath{\bm{\mu}}}
\newcommand{\edgeforce}{\ensuremath{\mu}}
\newcommand{\trailedgeforce}{\ensuremath{\edgeforce^{\text{t}}}}
\newcommand{\deviationedgeforce}{\ensuremath{\edgeforce^{\text{d}}}}

\newcommand{\edgelengths}{\ensuremath{\bm{\lambda}}}
\newcommand{\edgelength}{\ensuremath{\lambda}}
\newcommand{\trailedgelength}{\ensuremath{\edgelength^{\text{t}}}}
\newcommand{\deviationedgelength}{\ensuremath{\edgelength^{\text{d}}}}

\newcommand{\trailedgeplanes}{\ensuremath{\bm{\Phi}}}
\newcommand{\edgeplane}{\ensuremath{\bm{\phi}}}
\newcommand{\trailedgeplane}{\ensuremath{\edgeplane}}
\newcommand{\trailedgeplaneposition}{\ensuremath{\nodeposition^{\edgeplane_{i, j}}}}
\newcommand{\trailedgeplanenormal}{\ensuremath{\mathbf{n}^{\edgeplane_{i, j}}}}

\newcommand{\nodepositions}{\ensuremath{\mathbf{P}}}
\newcommand{\nodeposition}{\ensuremath{\mathbf{p}}}

\newcommand{\nodeloads}{\ensuremath{\mathbf{Q}}}
\newcommand{\nodeload}{\ensuremath{\mathbf{q}}}

\newcommand{\supportforces}{\ensuremath{\mathbf{R}}}
\newcommand{\supportforce}{\ensuremath{\mathbf{r}}}

\newcommand{\noderesidual}{\ensuremath{\mathbf{t}}}

\newcommand{\deviationforcevector}{\ensuremath{\mathbf{d}}}

\newcommand{\edgeloadpath}{\ensuremath{\varphi}}

\newcommand{\penaltyfunction}{\ensuremath{\mathcal{L}}}
\newcommand{\penaltyoutput}{\ensuremath{\penaltyfunction(\optimizationvariables)}}

\newcommand{\optimizationvariables}{\ensuremath{\mathbf{s}}}

\newcommand{\weight}{\ensuremath{w}}

\newcommand{\gradient}{\ensuremath{\nabla_{\optimizationvariables}\penaltyfunction}}
\newcommand{\gradientvalue}{\ensuremath{\nabla_{\optimizationvariables}\penaltyoutput}}

\newcommand{\convergencethreshold}{\ensuremath{\epsilon}}
\newcommand{\gradientthreshold}{\ensuremath{\kappa}}
\newcommand{\convergencecriteria}{\ensuremath{\penaltyfunction(\optimizationvariables)\leq\convergencethreshold \text{ or } \norm{\nabla\penaltyfunction(\optimizationvariables)}\leq\gradientthreshold}}

\newcommand{\constraint}{\ensuremath{g}}

\usepackage{natbib}

\usepackage{mathtools}
\usepackage{amsmath}
\usepackage{amssymb}
\usepackage{bm}
\let\norm\undefined 
\DeclarePairedDelimiter\norm{\lVert}{\rVert}

\usepackage{lipsum}
\usepackage{soul} 
\usepackage[dvipsnames]{xcolor}  
\usepackage{pdflscape} 
\usepackage{enumitem} 
\usepackage{ragged2e}
\usepackage[hyphens]{url} 
\usepackage[colorlinks=True, breaklinks=true]{hyperref} 
\hypersetup{allcolors=blue, pdfauthor=author}
\usepackage[nameinlink]{cleveref}
\usepackage{setspace}

\usepackage[linesnumbered,ruled,noend]{algorithm2e} 
\SetAlFnt{\normalsize}  
\SetInd{0.0em}{0.8em}  
\DontPrintSemicolon  
\SetSideCommentRight  
\SetKwComment{Comment}{$\triangleright$\ }{}

\usepackage{caption}
\usepackage{subcaption}
\usepackage{float} 
\usepackage[export]{adjustbox} 
\usepackage{stfloats} 

\usepackage{forest}
\usepackage{tikz}
\usetikzlibrary{shapes, arrows, calc, positioning, patterns}
\usetikzlibrary{arrows.meta}
\usetikzlibrary{backgrounds}
\newcommand\tikzmark[1]{\tikz[remember picture] \node (#1) {};}  
\definecolor{colororiginnode}{RGB}{153,150,255}
\definecolor{colorsupportnode}{RGB}{0,150,80}
\definecolor{colortensionedge}{RGB}{227,6,75}
\definecolor{colorcompressionedge}{RGB}{12,120,184}
\definecolor{colorauxtrail}{RGB}{255,155,15}

\usepackage{booktabs} 
\usepackage{multirow} 

\usepackage[frozencache,cachedir=.]{minted}

\journal{arXiv}
\begin{document}

\begin{frontmatter}

\title{Constrained Form-Finding of Tension-Compression Structures using Automatic Differentiation}

\author[1]{Rafael Pastrana\corref{cor1}}
\ead{arpastrana@princeton.edu}
\author[2]{Patrick Ole Ohlbrock}
\author[3]{Thomas Oberbichler}
\author[4]{Pierluigi D'Acunto}
\author[1]{Stefana Parascho}

\cortext[cor1]{Corresponding author}

\address[1]{CREATE Laboratory, School of Architecture, Princeton University, United States of America}
\address[2]{Chair of Structural Design, Institute of Technology in Architecture, ETH Zürich, Switzerland}
\address[3]{Chair of Structural Analysis, Technische Universität München, Germany}
\address[4]{Professorship of Structural Design, Technische Universität München, Germany}


\vspace{-10mm}
\begin{abstract}
This paper proposes a computational approach to form-find pin-jointed, bar structures subjected to combinations of tension and compression forces.
The generated equilibrium states can meet force and geometric constraints via gradient-based optimization.
We achieve this by extending the combinatorial equilibrium modeling (CEM) framework in three important ways. 
First, we introduce a new topological object, the auxiliary trail, to expand the range of structures that can be form-found with the framework. 
Then, we leverage automatic differentiation (AD) to obtain an exact value of the gradient of the sequential and iterative calculations of the CEM form-finding algorithm, instead of a numerical approximation. 
Finally, we encapsulate our research developments into an open-source design tool written in Python that is usable across different CAD platforms and operating systems.
After studying four different structures -- a self-stressed planar tensegrity, a tree canopy, a curved bridge, and a spiral staircase -- we demonstrate that our approach enables the solution of constrained form-finding problems on a diverse range of structures more efficiently than in previous work.
\end{abstract}


\begin{keyword}
form-finding \sep
optimization \sep
automatic differentiation \sep
design tool \sep
structural design \sep
combinatorial equilibrium modeling
\end{keyword}

\end{frontmatter}
\raggedbottom
\hyphenpenalty=1000



\section{Introduction}
\label{intro}

A \emph{form-finding} method generates the shape and the internal force state of a structure so that, given a design load case and a set of support conditions, the structure is in a state of \emph{static equilibrium} \cite{lewis_tensionstructures_2003a, bletzinger_fiftyyears_2011}.
Different numerical form-finding methods exist, but they fall into one of three categories: stiffness-matrix \cite{argyris_generalmethod_1974a, tabarrok_nonlinearanalysis_1992}, dynamic equilibrium \cite{barnes_formfinding_1999, kilian_particlespringsystems_2005, adriaenssens_findingform_2012} and geometric \cite{linkwitz_einigebemerkungen_1971, schek_forcedensity_1974, bletzinger_generalfinite_1999, nouribaranger_formfinding_2002, block_thrustnetwork_2007, pauletti_naturalforce_2008, dacunto_vectorbased3d_2019, hablicsek_algebraic3d_2019} approaches.
An in-depth review of this taxonomy is found in \cite{lewis_tensionstructures_2003a, veenendaal_overviewcomparison_2012}.
Developments in all categories have been propelled over the last decade by the development of multiple computational design tools
\cite{fivet_fullygeometric_2015, rippmann_funicularshell_2016, senatore_interactiverealtime_2015, lee_disjointedforce_2018, nejur_polyframeefficient_2021}.

In these form-finding approaches, a structure is often modeled as a discrete network of straight bars that are connected by pinned joints at the nodes.
The design load is transferred from node to node exclusively through axial forces in the bars and a state of static equilibrium is reached when the sum of forces incident to every node is zero.
Conceptually, the axial-dominant load-carrying mechanism of a structure in static equilibrium implies that it will require less material volume to withstand the applied design load \cite{michell_limitseconomy_1904}.

\subsection{The CEM framework}

The Combinatorial Equilibrium modeling (CEM) framework is a geometric form-finding method for spatial structures modeled as pin-jointed bar networks and subjected to combinations of tension and compression forces \cite{ohlbrock_combinatorialequilibrium_2016, ohlbrock_computeraidedapproach_2020, ohlbrock_combinatorialequilibrium_2020}.
Examples of such mixed systems are space frames, bridges, stadium roofs, multistory buildings and tensegrities.

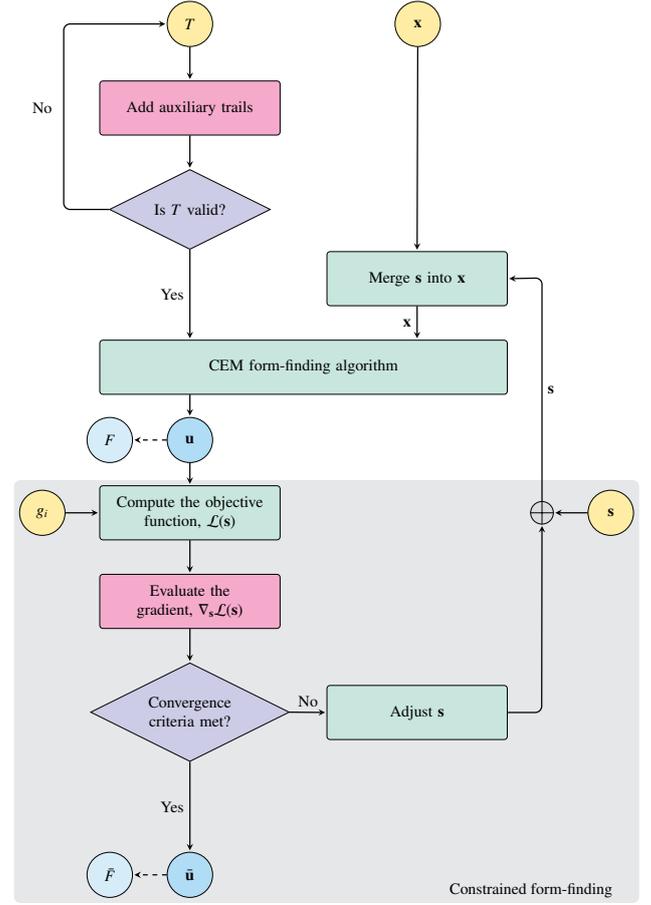
\begin{figure}[t!]  
  \centering
  \resizebox{\columnwidth}{!}{
    \tikzstyle{input} = [circle, minimum size=1cm, draw=black, fill=Goldenrod!50]  
\tikzstyle{output} = [circle, minimum size=1cm, draw=black, fill=Cerulean!30]
\tikzstyle{output_extra} = [circle, minimum size=1cm, draw=black, fill=Cerulean!15]
\tikzstyle{or} = [circle, minimum size=0.5cm, draw=black, fill=Gray!30]

\tikzstyle{process} = [rectangle, inner sep=-0.1ex, minimum height=1.2cm, text centered, text width=4cm, draw=black, rounded corners=2, fill=SeaGreen!30]
\tikzstyle{half_process} = [rectangle, inner sep=-0.1ex, minimum height=1.2cm, text centered, text width=2cm, draw=black, rounded corners=2, fill=SeaGreen!30]
\tikzstyle{double_process} = [rectangle, inner sep=-0.1ex, minimum height=1.2cm, text centered, text width=9cm, draw=black, rounded corners=2, fill=SeaGreen!30]  

\tikzstyle{decision} = [diamond, aspect=2.0, inner sep=-1ex,  text centered, text width=4cm,  draw=black, fill=Periwinkle!30]

\tikzstyle{arrow} = [thick, ->, >=stealth, shorten >=1pt, rounded corners]
\tikzstyle{arrow_dashed} = [thick, dashed, ->, >=stealth, shorten >=1pt, rounded corners]

\begin{tikzpicture}

\node (topology) [input]  at (0,0) {$\topology$};

\node(parameters) [input, right=4.0cm of topology] {$\designparameters$};

\node (auxiliary) [process, below=0.75cm of topology, fill=Rhodamine!50] {Add auxiliary trails};

\node (validity) [decision, below=0.75cm of auxiliary] {Is $\topology$ valid?};
\node (invalid) [circle, minimum size=1cm, left=0.75cm of auxiliary.west] {No};

\node (equilibrium) [double_process, below=2cm of validity, xshift=2.5cm] {CEM form-finding algorithm};

\node (par_update) [process, above=0.75cm of equilibrium, xshift=2.5cm] {Merge $\optimizationvariables$ into $\designparameters$};

\node (form) [output, below=0.5cm of equilibrium, xshift=-2.5cm] {$\equilibriumattributes$};
\node (diagram) [output_extra, left=0.75cm of form] {$\form$};
    
\node (cost) [process, below=0.5cm of form] {Compute the objective function,  $\penaltyfunction(\optimizationvariables)$};

\node (gradient) [process, below=0.75cm of cost, fill=Rhodamine!50] {Evaluate the gradient, $\gradient(\optimizationvariables)$};

\node (convergence) [decision, below=0.75cm of gradient] {Convergence\\criteria met?};

\node (optimization) [process, below=1.25cm of gradient, xshift=5cm] {Adjust $\optimizationvariables$};  

\node(par_or2) [or, right=5.5cm of cost] {};
\draw (par_or2.north) -- (par_or2.south) (par_or2.west) -- (par_or2.east);

\node (form_opt) [output, below=2cm of convergence] {$\constrainedequilibriumattributes$};

\node (diagram_opt) [output_extra, left=0.75cm of form_opt] {$\constrainedform$};

\node(constraints) [input, left=0.75cm of cost.west] {$\constraint_i$};

\node(par_opt) [input, right=0.75cm of par_or2.east] {$\optimizationvariables$};

\draw[arrow] (topology.south) -- (auxiliary.north)
    node[midway, left]{};

\draw[arrow] (parameters.south) -- (par_update.north) node[midway, left]{};

\draw[arrow] (par_update.south) -- ([xshift=2.5cm]equilibrium.north) node[midway, left]{$\designparameters$};

\draw [arrow] (par_opt.west) -- (par_or2.east);

\draw[arrow] (auxiliary.south) -- (validity.north) node[midway, left]{};

\draw[arrow] (validity.south) -- ([xshift=-2.5cm]equilibrium.north) node[midway, left]{Yes};

\draw[arrow] ([xshift=-2.5cm]equilibrium.south) -- (form.north) node[midway, left]{};

\draw[arrow] (form.south) -- (cost.north) node[midway, left]{};

\draw[arrow_dashed] (form.west) -- (diagram.east) node[midway, left]{};

\draw[arrow] (cost.south) -- (gradient.north) node[midway, left]{};

\draw[arrow] (gradient.south) -- (convergence.north) node[midway, left]{};

\draw[arrow] (convergence.south) -- (form_opt.north) node[midway, left]{Yes};

\draw[arrow] (convergence.east) -- (optimization.west) node[midway, above]{No};

\draw[arrow_dashed] (form_opt.west) -- (diagram_opt.east) node[midway, left]{};
    
\draw[arrow] (constraints.east) -- (cost.west);

\draw [arrow] (validity.west) -- ++(-1.0,0) |- (topology.west);

\draw [arrow] (optimization.east) -- ++(0.75,0) -- (par_or2.south) node (X) [midway, right]{};

\draw [arrow] (par_or2.north) -- ++(0,4.95) node (X) [midway, right]{$\optimizationvariables$} -- (par_update.east) ;

\begin{scope}[on background layer]
    \node (fitconstrained) [rounded corners, fill=Gray!20, fit=(cost) (form_opt) (constraints) (par_opt), label={[rotate=0, anchor=south, xshift=4.5cm] below: Constrained form-finding}] {};
\end{scope}

\end{tikzpicture}}
    \caption{Overview of the Combinatorial Equilibrium Modeling (CEM) framework and our extensions.
    The inputs to the framework are a topology diagram $\topology$ and the design parameters $\designparameters$.
    The CEM form-finding algorithm calculates a state of static equilibrium $\equilibriumattributes$ from which a form diagram $\form$ can be optionally constructed.    
    To find a constrained equilibrium state $\constrainedequilibriumattributes$ that best satisfies force and geometric constraints $\constraint_i$, the CEM framework minimizes an objective function $\penaltyoutput$ by iteratively adjusting the optimization parameters $\optimizationvariables$ using the gradient $\gradient(\optimizationvariables)$ until the convergence criteria $\convergencecriteria$ is reached.
    Two of the extensions we make in this paper are highlighted in pink.
    Auxiliary trails simplify the construction of a larger variety of valid topology diagrams $\topology$.
    Reverse-mode automatic differentiation computes an exact value of $\gradient(\optimizationvariables)$ thus allowing for more efficient and stable solutions to constrained form-finding problems.
    }
    \label{fig:framework}
\end{figure}

The framework consists of two operative parts: the CEM form-finding algorithm and an optimization-based constrained form-finding solver.
Moreover, this framework represents a structure with three diagrams: A topology diagram $\topology$ describes its internal connectivity and internal tension-compression state.
Meanwhile, a form diagram $\form$ and a force diagram $\force$ display the geometric and force attributes of the calculated state of static equilibrium.
Figure \ref{fig:framework} presents a graphical overview of how these components interact and Section \ref{cemf} provides a thorough review of their underpinnings.

A distinctive feature of the CEM form-finding algorithm is that equilibrium is computed sequentially and iteratively, unlike other geometric form-finding methods \cite{linkwitz_einigebemerkungen_1971, schek_forcedensity_1974,bletzinger_generalfinite_1999,bahr_formfinding_2017}.
Nevertheless, this computation approach is precisely what allows the CEM form-finding algorithm to ensure the generation of a static equilibrium state for a mixed tension-compression structure as long as an input topology diagram $\topology$ fulfills the requirements listed in Section \ref{rules}. 

The CEM algorithm allows designers to explore different equilibrium states for a fixed diagram $\topology$ by manipulating a portion of the nodal positions, and a subset of the bar forces and lengths of a structure (see Section \ref{parameters}).
However, realistic structural design scenarios often pose geometrical and force constraints a priori where it is more relevant to find a specific equilibrium state that best satisfies them.
Examples of such constraints include fitting a target shape \cite{panozzo_designingunreinforced_2013, tamai_advancedapplication_2013}, restraining bar forces and lengths \cite{zhang_adaptiveforce_2006, allen_formforces_2009, miki_extendedforce_2010}, and controlling the reaction forces at the supports of a structure \cite{malerba_extendedforce_2012, quagliaroli_flexiblebridge_2013, lee_disjointedforce_2018}.
The challenge is that while the design constraints can be readily enumerated, it is often not straightforward to discern what combination of input design parameters is conducive to the envisioned result. 

One way to solve such a constrained form-finding problem is to manually tweak the input design parameters until the required constraints are met, one by one. This can quickly become a cumbersome process.
Instead, the CEM framework form-finds spatial structures subjected to geometrical and force constraints following an automatic approach: constraints are aggregated into a single objective function and a computer assists the designer in calculating the values of the design parameters that minimize it via gradient-based optimization \cite{ohlbrock_computeraidedapproach_2020, ohlbrock_constraintdrivendesign_2017}. 

\subsection{Limitations of the CEM framework}\label{limitations}

The CEM framework as presented in \cite{ohlbrock_combinatorialequilibrium_2016, ohlbrock_constraintdrivendesign_2017, ohlbrock_computeraidedapproach_2020} faces two limitations. 
One of them is related to its topological modeling flexibility and the other to its computational performance when solving a constrained form-finding problem.

\subsubsection{Strict topological modeling rules}\label{rules}

Every topology diagram $\topology$ must fulfill two requirements in order to be considered a valid input to the CEM form-finding algorithm:
        
\begin{enumerate}
  \itemsep 0em
  \item Every node $\node$ needs to be part of only one trail $\trail$ (see Section \ref{trails}).
  \item Every trail $\trail$ must have only one support assigned to its last node.
\end{enumerate}

Abiding by these two rules restricts the type of structures that the CEM algorithm can form-find.
Adequately constructing a valid topological diagram can become a daunting task without a sound knowledge of the CEM theoretical background, especially for structures that do not have a clear load-transfer hierarchy.

\begin{figure*}[t]
    \hspace{-2.0cm}
        \centering
        \begin{subfigure}[b]{0.37\textwidth}
            \centering
            \includegraphics[width=\textwidth,trim={0 3mm 0 3mm},clip]{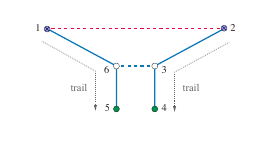}
            \caption{Two-sided cantilever structure}
            \label{fig:tree_2d}
        \end{subfigure}
        \hspace{-1.0cm}
        \begin{subfigure}[b]{0.37\textwidth}
            \centering
            \includegraphics[width=\textwidth,trim={0 3mm 0 3mm},clip]{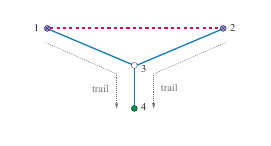}
            \caption{Branching structure}
            \label{fig:tree_2d_wrong}
        \end{subfigure}
        \hspace{-1.0cm}
        \begin{subfigure}[b]{0.37\textwidth}
            \centering
            \includegraphics[width=\textwidth,trim={0 3mm 0 3mm},clip]{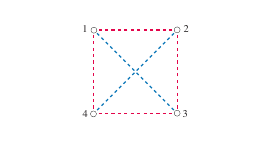}
            \caption{Self-stressed tensegrity}
            \label{fig:self_stressed_wrong}
        \end{subfigure}
        \hspace{-2.0cm}
    
    \caption{Topology diagrams $\topology$ that correspond to three different structural systems. Note that diagrams \ref{fig:tree_2d_wrong} and \ref{fig:self_stressed_wrong} do not meet the CEM topology requirements listed in Section \ref{limitations}. The former shows nodes 3 and 4 as members of two different trails, $\trail_1$ and $\trail_2$, $\omega_2$, whereas the latter defines neither trails nor support nodes.}
    \label{fig:topological_diagrams}
\end{figure*}

Figure \ref{fig:tree_2d} shows the topological diagram of a structure in which the two topological rules are satisfied. 
In contrast, the diagram in Figure \ref{fig:tree_2d_wrong} violates the first rule because the two proposed trails $\trail_1 = \{1, 3, 4\}$ and $\trail_2 = \{2, 3, 4\}$ share nodes 3 and 4. 
Figure \ref{fig:self_stressed_wrong} depicts a planar, self-stressed tensegrity structure which by definition has no supports and consequently infringes rule number two.

\subsubsection{Approximate gradient computation}

To solve a constrained form-finding problem, the CEM framework has approximated the gradient of the objective function to minimize via finite differences (FD) \cite{ohlbrock_constraintdrivendesign_2017,ohlbrock_computeraidedapproach_2020}.
This is in stark contrast to other geometric form-finding precedents where the analytical equations to calculate an exact gradient have been published \cite{schek_forcedensity_1974, panozzo_designingunreinforced_2013, quagliaroli_flexiblebridge_2013, tamai_advancedapplication_2013, takahashi_advancedform_2018, cuvilliers_gradientbasedoptimization_2016}.
FD circumvents the derivation problems we discuss in Section \ref{autograd} as it does not require the calculation of analytical derivatives of the CEM form-finding algorithm to obtain a gradient estimate.

Using FD poses a number of challenges nonetheless.
FD requires choosing an adequate step size $h$ to compute the gradient approximation \cite{nocedal_numericaloptimization_2006}. 
If the resulting step size $h$ is too large, the gradient approximation can be inaccurate, whereas if it is too small it can lead to significant round-off errors due to numerical underflow.
Moreover, calculating gradients with FD is computationally taxing as the objective function has to be evaluated at least once per every optimization parameter input \cite{nocedal_numericaloptimization_2006, haase_optimalsizing_2001}.
This can be detrimental to an interactive exploration of static equilibrium states for a structure, particularly if the number of optimization parameters is large.

\subsection{Automatic differentiation}

\emph{Automatic differentiation} (AD), also known as algorithmic differentiation, comprises a set of techniques that evaluate the derivative of a differentiable function that is expressed algorithmically \cite{corliss_automaticdifferentiation_2013}.
In contrast to FD, derivatives obtained with AD are exact up to computer precision and do not require the specification of a step size $h$ to be computed \cite{nocedal_numericaloptimization_2006}.
Unlike symbolic differentiation, derivatives can be evaluated with AD through control flow statements, such as if-else clauses, loops and recursion \cite{corliss_automaticdifferentiation_2013}.

One of the prominent applications of AD today is in training machine learning models via backpropagation, in particular neural architectures that learn via gradient-based optimization \cite{baydin_automaticdifferentiation_2018}.
Other precedents of AD applied to various engineering problems are the sizing of the frame of an injection molding machine \cite{haase_optimalsizing_2001}, the shape optimization of a supersonic aircraft \cite{unger_useautomatic_1994}, and the weight minimization of steel frames under seismic loads \cite{cho_multilevelmultiobjective_2000}.
In addition to its ease of use, AD is also characterized by its high performance.
For example, AD has been used to automatically derive complex isogeometric elements from an energy functional \cite{oberbichler_efficientcomputation_2021}.

In the context of form-finding, Cuvilliers recently proposed to use AD in form-finding framework for structures subjected to geometric constraints \cite{cuvilliers_constrainedgeometry_2020}.
Unlike their work, we use the CEM form-finding algorithm and not the \emph{Force Density Method} \cite{linkwitz_einigebemerkungen_1971, schek_forcedensity_1974} as the equilibrium state calculator.
Furthermore, they focus on compression-only shells while we study various types of mixed tension-compression structures.

\subsection{Outline and contribution}\label{outline}

Table \ref{table:symbology} lists the symbols we use to display topology, form and constrained form diagrams, $\topology$, $\form$ and $\constrainedform$, respectively.

This paper is organized in six sections.

In Section \ref{cemf} we present the theoretical concepts that underpin the current state of the CEM framework.
We review the steps followed to sequentially and iteratively compute a state of static equilibrium from an algorithmic perspective, and discuss the mathematical formulation of the objective function that is minimized to solve a constrained form-finding problem.

Section \ref{extensions} develops the extensions we make to the CEM framework to overcome the limitations we outlined in Section \ref{limitations}, which constitute the core of our contribution.
We first introduce a new topological helper object, the auxiliary trail.
We show next how we leverage AD to evaluate an exact gradient of the CEM form-finding algorithm and guide the reader through this process with a simple constrained form-finding example.
We also present a new standalone design tool called \texttt{compas\_cem} that encapsulates the two above-mentioned extensions.

In Section \ref{experiments}, we benchmark and validate the extended CEM framework by studying multiple constrained form-finding problems on three different types of structures: a self-stressed tensegrity, a tree canopy and a curved suspension bridge subjected to torsional loads. 
The applicability of our work in practical structural design problems is showcased in the case study of a spiral staircase in Section \ref{casestudy}.
For reproducibility, we share the data and the code we use to solve all the constrained form-finding problems we discuss in Sections \ref{experiments} and \ref{casestudy} in the open-access repository in \cite{cemad-cad}.

The paper concludes in Section \ref{conclusion} with a discussion of our experimental findings, the limitations of our approach, its potential relation to other geometric form-finding methods, and future research directions.

\begin{table}[!b]
    \centering
    \tikzstyle{mynode} = [circle, draw=black, line width=.7pt, inner sep=1mm] 

\begin{tabular}{@{}p{17mm}cl@{}}
    \toprule

    \centering Diagram&
    Symbol&
    Description\\
    
    \midrule

    \multirow{9}{17mm}{\centering$\topology$}&
    \tikzmark{node}&
    Node\\

    &
    \tikzmark{originnode}&
    Origin node\\

    & 
    \tikzmark{supportnode}&
    Support node\\

    \cmidrule(lr){2-3}

    &
    \tikzmark{load}&
    Load\\

    \cmidrule(lr){2-3}

    & 
    \tikz{\draw[color=colortensionedge, line width=1.2pt] (0,0)--(0.9,0);}&
    Trail edge in tension\\

    & 
    \tikz{\draw[color=colorcompressionedge, line width=1.2pt] (0,0)--(0.9,0);}&
    Trail edge in compression\\

    & 
    \tikz{\draw[color=colortensionedge, line width=1.2pt, dash pattern=on 3pt off 1.5pt] (0,0)--(0.9,0);}&
    Deviation edge in tension\\

    & 
    \tikz{\draw[color=colorcompressionedge, line width=1.2pt, dash pattern=on 3pt off 1.5pt] (0,0)--(0.9,0);}&
    Deviation edge in compression\\

    & 
    \tikz{\draw[color=colorauxtrail, line width=1.2pt] (0,0)--(0.9,0);}&
    Auxiliary trail edge\\

    \midrule[0.5pt]

    \multirow{4}{17mm}{\centering$\form$,\;$\constrainedform$}&
    \tikzmark{formnode}&
    Node\\
    
    \cmidrule(lr){2-3}
    
    &
    \tikz{\draw[-{latex[length=5mm]}, color=colorsupportnode, line width=1pt] (0,0)--(0.9,0);}&
    Load or reaction force vector\\
    
    \cmidrule(lr){2-3}

    & 
    \tikz{\draw[color=colortensionedge, line width=1.2pt] (0,0)--(0.9,0);}&
    Edge in tension\\

    & 
    \tikz{\draw[color=colorcompressionedge, line width=1.2pt] (0,0)--(0.9,0);}&
    Edge in compression\\

    \bottomrule

\end{tabular}

\tikz[remember picture,overlay]{\node[mynode] at (node.center) {}}
\tikz[remember picture,overlay]{\node[mynode, fill=colororiginnode] at (originnode.center) {}}
\tikz[remember picture,overlay]{\node[mynode, fill=colorsupportnode] at (supportnode.center) {}}
\tikz[remember picture,overlay]{\node[mynode, fill=white, draw=white, fill opacity=0.0, draw opacity = 0.0] (loaded) at (load.center) {}; 
\draw[line width=1.0pt] (loaded.north east) -- (loaded.south west) (loaded.north west) -- (loaded.south east);}
\tikz[remember picture,overlay]{\node[mynode] at (formnode.center) {}}
    \caption{Symbols that describe the elements of topology ($\topology$), form ($\form$) and constrained form ($\constrainedform$) diagrams in the extended CEM framework.
    Refer to Section \ref{topology_diagram} for element definitions.}
    \label{table:symbology}
\end{table}

\section{Theoretical background}
\label{cemf}

We review how the Combinatorial Equilibrium Modeling (CEM) framework works.
This is relevant to guide the discussion in subsequent sections.
The CEM framework was introduced in \cite{ohlbrock_combinatorialequilibrium_2016} and further developed in \cite{ohlbrock_computeraidedapproach_2020, ohlbrock_combinatorialequilibrium_2020}.
It consists of two operative parts: (i) a form-finding algorithm that builds equilibrium sequentially and iteratively (Section \ref{fofin}), and (ii) a constrained form-finding routine that utilizes an optimization solver (Section \ref{constrained_fofin}).
The goal of the former is to generate a numerical \emph{state of static equilibrium}, $\equilibriumattributes$.
That of the latter is to produce a \emph{constrained state of static equilibrium} $\constrainedequilibriumattributes$ that has been restricted by a set of geometric and force constraints.
Regardless, there are two necessary inputs to calculate these states: a valid topology diagram $\topology$ and a vector of design parameters $\designparameters$.

\subsection{Topology diagram}\label{topology_diagram}

A \emph{topology diagram} $\topology$ is an undirected graph of $N$ nodes $\nodes$ connected by $M$ edges $\edges$.
It captures the internal connectivity of a structure modeled as a pin-jointed network of straight bars.

Every edge $\edge_{i, j}$ connecting two nodes $\node_i, \node_j$ must be labeled as either a \emph{trail edge} $\trailedge_{i, j}$ or a \emph{deviation edge}, $\deviationedge_{i, j}$. 
Therefore, the total number of trail edges $R$ and the total number of deviation edges $D$ in $\topology$ must add up add up to $M$, i.e. $R\,+\,D\,=\,M$.
Trail edges outline the trails of a structure (Section \ref{trails}).
These edges determine the primary paths that the loads applied to a structure follow towards the supports.
Deviation edges connect nodes on different trails to redirect the load trajectories determined by the trails.

The entries in the adjacency matrix $\internalforcestates\in\{-1,0,1\}$ of the topology diagram define the expected internal force state $\internalforcestate_{i, j}$ of the bars in the structure \cite{ohlbrock_computeraidedapproach_2020}.
If $\internalforcestates[i, j]=\internalforcestate_{i, j}=-1$, the corresponding edge $\edge_{i, j}$ is in compression. Conversely, if $\internalforcestate_{i, j}=1$, then $\edge_{i, j}$ is in tension.
A topology diagram furthermore prescribes the subset of size $L$ with the nodes $\supports$ where a support is assigned, $\supports \subset \nodes$. 

\subsubsection{Trails}\label{trails}

Trails are critical to evaluate the validity of a topology diagram $\topology$, as discussed in Section \ref{rules}.
A \emph{trail} $\trail$ is an ordered set of nodes that are linked exclusively by trail edges, $\trail = \{ \nodeorigin, ..., \nodesupport \}$. 
The first node in a trail $\nodeorigin$ is referred to as an \emph{origin node}. 
The last node $\nodesupport$ must have a support assigned, $\nodesupport \in \supports$, and it is thus referred to as a \emph{support node}.
A trail must contain at least two nodes. The set of all trails in a topology diagram is denoted $\trails$.
There must be as many trails as there are support nodes, $|\supports|=|\trails|$. 
Trails need not have the same number of nodes.

\subsubsection{Sequences}\label{sequences}

Once a trail $\trail$ is constructed, the nodes within are sorted based on how distant they are to the origin node $\nodeorigin$ in the trail.
For every node $\node$, this topological distance is defined as the number of intermediate trail edges $\trailedge$ plus one between $\node$ and $\nodeorigin$.
Nodes that are equally distant to their corresponding $\nodeorigin$ belong to the same \emph{sequence}, $\sequence$.
While the first sequence $\sequence=1$ groups all the origin nodes $\nodeorigin$ in all possible trails $\Omega$, the last sequence$\sequence^\text{last}$ contains the support nodes of the trails with the highest number of nodes.
The list of sequences $\sequences$ is an ordered set of consecutive integers between $\sequence=1$ and $\sequence^\text{last}$.

\subsection{Design parameters}\label{parameters}

The vector of design parameters $\designparameters$ prescribes an immutable portion of the state of static equilibrium $\equilibriumattributes$ that is calculated by the CEM form-finding algorithm.
It concatenates:

\begin{itemize}
    \itemsep 0em
    \item A vector $\edgeforces \in \mathbb{R}_{+}^{M}$ with the absolute magnitude of the internal force $\deviationedgeforce_{i, j}$ of every deviation edge $\deviationedge_{i, j}$.
    \item A vector $\edgelengths \in \mathbb{R}_{+}^{M}$ with the length $\trailedgelength_{i, j}$ of each trail edge $\trailedge_{i, j}$.
    Alternatively, a matrix $\trailedgeplanes\in\mathbb{R}^{R \times 6}$ with an intersection plane $\trailedgeplane_{i, j}\in\mathbb{R}^6$ per trail edge $\trailedge_{i, j}$ to implicitly compute $\trailedgelength_{i, j}$.
    \item A matrix $\nodepositions \in \mathbb{R}^{N\times3}$ with the position $\nodeposition^{\text{o}}\,\in\,\mathbb{R}^3$ of every origin node $\nodeorigin$
    \item A matrix $\nodeloads \in \mathbb{R}^{N\times3}$ with the load vectors $\nodeload\,\in\,\mathbb{R}^3$ applied to the nodes $\node$. Only one load vector per node is permitted. If the modeled structure is self-stressed, all the entries in $\nodeloads$ are null vectors.
\end{itemize}

\subsection{Form-finding algorithm}\label{fofin}

The CEM form-finding algorithm completes the attributes in $\equilibriumattributes$ following Algorithm \ref{fofin_algo}.
The numerical outputs of the algorithm consist of:

\begin{itemize}
    \itemsep 0em 
    \item The absolute magnitude of the internal force $\trailedgeforce_{i, j} \in \edgeforces$ of every trail edge $\trailedge_{i, j}$.
    \item The length $\deviationedgelength_{i, j} \in \edgelengths$ of every deviation edge $\deviationedge_{i, j}$.
    \item The position $\nodeposition_i \in \nodepositions$ of every non-origin node, $\node_i \neq \nodeorigin_i$.
    \item A matrix $\supportforces \in \mathbb{R}^{L\times3}$ with the reaction force vector $\supportforce_i\in \mathbb{R}^{3}$ incident to every support node, $\nodesupport_i \in \supports$.
\end{itemize}

These outputs are concatenated into a single vector.
Once $\equilibriumattributes$ is complete, the form diagram $\form$ and the force diagram $\force$ of the structure can be built using vector-based graphic statics \cite{dacunto_vectorbased3d_2019} to visualize the resulting equilibrium state of the structure. 

\begin{algorithm}[!tb]
  \caption{The CEM form-finding algorithm}
  \label{fofin_algo}
  \setstretch{1.05}
  \SetKwInOut{Input}{Input}
\SetKwInOut{Output}{Output}

\Input{Topology diagram, $\topology$\newline
    Sequences, $\sequences$\newline
    Trails, $\trails$\newline
    Design parameters, $\designparameters$\newline
    Maximum \# of equilibrium iterations, $\iterationsmax$\newline
    Minimum distance threshold, $\energythreshold$}
\Output{State of static equilibrium, $\equilibriumattributes$}

$ \nodes, \edges, \supports, \internalforcestates, \leftarrow \topology$\;
$ \edgeforces, \edgelengths, \trailedgeplanes, \nodepositions, \nodeloads \leftarrow \designparameters$\;
iteration, $\iteration \leftarrow 1$\;
distance, $\energy \leftarrow \infty$\;

\While{$\iteration \leq \iterationsmax \textbf{\upshape{ or }} \energy \geq \energythreshold$}
{
    \For{\text{\upshape{sequence \textbf{in} sequences, }}$\sequence \in \sequences$}
    {

    {\For{$\text{\upshape{trail \textbf{in} trails, }} \trail\in\trails$}
        { 
            $\node_i \leftarrow \text{\upshape{NodeInTrailAtSequence($\trail, \sequence$)}}$\;

            \If{$\text{\upshape{node }}\node_i \text{\upshape{ exists}}$}
            {
            
            \If{$\text{\upshape{first sequence, }} \sequence=1$}
            {$\noderesidual_h \leftarrow \mathbf{0}$\;
            $\nodeposition_i\leftarrow\text{PositionOriginNodeInTrail($\trail$)}$}

            $\deviationforcevector_i \leftarrow$ \text{DeviationEdgesVector($\node_i$)}\Comment*[r]{Eq.\ref{deviation_vector}}
            
            $\noderesidual_i \leftarrow \text{ResidualForceVector($\noderesidual_h, \deviationforcevector_i, \nodeload_i$)}$\Comment*[r]{Eq.\ref{node_residual}}
            
            \eIf{$\text{\upshape{node $\node_i$ is \textbf{not} a support, }} \node_i$$ \notin \supports$}
            {

            \If{$\text{\upshape{plane }} \trailedgeplane_{i, j} \text{\upshape{ exists}}$}
            {$\trailedgelength_{i, j} \leftarrow \text{PlaneIntersection($\nodeposition_i,\noderesidual_i,\trailedgeplane_{i, j}$)}$\Comment*[r]{Eq.\ref{eq:plane_length}}}
            
            $\nodeposition_j \leftarrow \text{\upshape{NodePosition($\nodeposition_i,\noderesidual_i, \internalforcestate_{i, j},\trailedgelength_{i, j}$)}}$\Comment*[r]{Eq.\ref{node_xyz}}

            $\trailedgeforce_{i, j} \leftarrow \text{\upshape{TrailEdgeForce($\noderesidual_i$)}}$\Comment*[r]{Eq.\ref{force_trail}}
            $\noderesidual_h \leftarrow \noderesidual_i$\;
            $\nodeposition_i \leftarrow \nodeposition_j$\;
            }
            {
            $\supportforce_i \leftarrow \noderesidual_i$\;
            }
            }
        }
    }
    }
    \For{\text{\upshape{deviation edge \textbf{in} edges, }}$\deviationedge_{i, j} \in \edges$}
    {
    $\deviationedgelength_{i, j}\leftarrow\text{\upshape{DeviationEdgeLength($\nodeposition_i, \nodeposition_j$)}}$\Comment*[r]{Eq.\ref{length_deviation}}
    }
    
    \If{$\text{\upshape{\textbf{not} first iteration, }}\iteration>1$}{

    $\energy\leftarrow\text{\upshape{NodeDistances($\nodepositions^{(\iteration)}, \nodepositions^{(\iteration-1)}$)}}$\Comment*[r]{Eq.\ref{equilibrium_energy}}
    }
    
    $\iteration \leftarrow \iteration+1$\;
}
$\equilibriumattributes \leftarrow \edgeforces, \edgelengths, \nodepositions, \nodeloads, \supportforces$\;

\end{algorithm}

\subsubsection{Sequential equilibrium}
\label{sequential}

Static equilibrium is calculated at the nodes of the diagram $\topology$ one sequence at a time \cite{ohlbrock_combinatorialequilibrium_2016, ohlbrock_computeraidedapproach_2020}.
The form-finding process starts off by computing equilibrium at the nodes at the first sequence $\sequence=1$ and continues to the next sequence $\sequence+1$ until the last one, $\sequence^\text{last}$, is reached.
The calculation ends for every trail $\trail$ when the node at the current sequence $\sequence$ is a support node.

Except for the nodes at the first sequence $\sequence=1$, a state of static equilibrium at node $\node_j$ at sequence $\sequence\neq1$ is subordinated to the equilibrium state of the nodes preceding it.
Let the triplet of nodes $\node_h$, $\node_i$ and $\node_j$ be three consecutive nodes along a sequence-ordered trail $\trail$.
Let $\node_i$ be the node on the trail at sequence $\sequence$, $\node_h$ the node at the previous sequence $\sequence-1$ and $\node_j$ the node at the next sequence $\sequence+1$.
The calculation of static equilibrium at node $\node_i$ outputs the position $\nodeposition_j$ of the next node $\node_j$ and the force magnitude $\trailedgeforce_{i, j}$ of the trail edge $\trailedge_{i, j}$ connecting them.
The position $\nodeposition_j$ is described as

\begin{equation}\label{node_xyz}
  \nodeposition_j = \nodeposition_i + \internalforcestate_{i, j}\, \trailedgelength_{i, j}\,\frac{\noderesidual_i}{\norm{\noderesidual_i}}
\end{equation}

\noindent
where $\nodeposition_i$ is the position of node $\node_i$, $\internalforcestate_{i, j}$ and $\trailedgelength_{i, j}$ are the internal force state ($-1$ for compression, $+1$ for tension) and the length of $\trailedge_{i, j}$, respectively; and $\noderesidual_i$ is the residual force vector incident to node $\node_i$.

If a plane $\trailedgeplane_{i, j}$ is supplied instead of a specific trail edge length $\trailedgelength_{i, j}$, then the absolute magnitude of $\trailedgelength_{i, j}$ is computed by intersecting the line of action of the vector $\noderesidual_i$ onto $\trailedgeplane_{i, j}$ \cite{ohlbrock_combinatorialequilibrium_2020}:

\begin{equation}
  \label{eq:plane_length}
  \trailedgelength_{i, j} =
  \left|
  \frac{\trailedgeplanenormal \cdot (\trailedgeplaneposition - \nodeposition_i)}{\trailedgeplanenormal \cdot (\noderesidual_i / \norm{\noderesidual_i})}
  \right|
\end{equation}

\noindent
where $\trailedgeplaneposition\in\mathbb{R}^3$ is the base point and $\trailedgeplanenormal\in\mathbb{R}^3$ the vector normal that define the intersection plane $\trailedgeplane_{i, j}$.

To estimate vector $\noderesidual_i$, all the forces acting on $\node_i$ are summed:

\begin{equation}
\label{node_residual}
  \noderesidual_{i} = \noderesidual_{h} - \deviationforcevector_{i} - \nodeload_{i}
  \quad \textrm{where} \quad 
  \noderesidual_{h} =
  \begin{cases}
    \mathbf{0} & \text{if $\sequence=1$}\\
    \noderesidual_{i}^{(k-1)} & \text{otherwise}
  \end{cases}
\end{equation}

\noindent where $\noderesidual_h$ is the residual force vector at preceding node $\node_h$.
The vector $\nodeload_i$ denotes the load applied to node $\node_i$, if any, and $\deviationforcevector_i$ corresponds to the resultant force vector generated by all the deviation edges $\deviationedge_{i, m}$ connected to $\mathbf{v}_i$:

\begin{equation}\label{deviation_vector}
  \deviationforcevector_i = \sum_{m}\internalforcestate_{i, m}\,\deviationedgeforce_{i, m}\,\frac{\nodeposition_i - \nodeposition_m}{\norm{\nodeposition_i - \nodeposition_m}}
\end{equation}

The terms $\internalforcestate_{i, m}$, $\deviationedgeforce_{i, m}$ and $\nodeposition_m$ encode the force state, force magnitude and the position of the node $\node_m$ that is connected to $\node_i$ by the deviation edge $\deviationedge_{i, m}$, respectively.
If no deviation edges are connected to $\mathbf{v}_i$, then $\deviationforcevector_i=\mathbf{0}$.

To maintain the equilibrium of forces at $\node_i$, the residual vector $\noderesidual_i$ is taken by the trail edge $\trailedge_{i, j}$ such that the vector formed between positions $\nodeposition_i$ and $\nodeposition_j$ point in the same direction as $\noderesidual_i$, and the absolute magnitude of the force passing through the edge $\trailedgeforce_{i, j}$ is equal to the norm of $\noderesidual_i$:

\begin{equation}
\label{force_trail}
    \trailedgeforce_{i, j} = \norm{\noderesidual_i}
\end{equation}

If $\node_j$ is a support node the residual vector $\noderesidual_i$ is parsed as the reaction force vector incident to the support node $\nodesupport_i$, $\supportforce_i = \noderesidual_i$. 
The length of any deviation edge $\deviationedge_{i, j}$ is lastly calculated as the norm of the distance vector between the positions of the two nodes it links:

\begin{equation}\label{length_deviation}
    \deviationedgelength_{i, j} = \norm{\nodeposition_j - \nodeposition_b}
\end{equation}

\subsubsection{Iterative equilibrium}\label{iterative}

The process described in Section \ref{sequential} must be run iteratively whenever (i) form-dependent load cases like wind or self-weight are applied; or (ii) deviation edges $\deviationedge_{i, j}$ that connect any two nodes $\node_i, \node_j$ that do not belong to the same sequence $\sequence$ exist (also called \emph{indirect deviation edges}) \cite{ohlbrock_computeraidedapproach_2020}. 
The termination conditions here are to exhaust a maximum number of iterations $\iterationsmax$ or to reach a minimum distance threshold $\energythreshold$ close to zero, such that $\energy \leq \energythreshold$:

\begin{equation}\label{equilibrium_energy}
    \energy = \sum_{i}^{\nodes} \norm[\big]{\nodeposition_i^{(\iteration)} - \nodeposition_i^{(\iteration - 1)}}
\end{equation}

The distance $\energy$ measures the cumulative displacement of the position $\nodeposition_i$ of every node $\node_i$ at iteration $\iteration$ in relation to the previous one.
The value of $\energy$ can be normalized by dividing it by $N$ to make it independent of the total number of nodes in the structure.
If indirect deviation edges exist, their contribution to $\deviationforcevector_i$ in Equation \ref{deviation_vector} is set to $\mathbf{0}$ during the first iteration, $\iteration=1$ \cite{ohlbrock_computeraidedapproach_2020}.

\subsection{Constrained form-finding}
\label{constrained_fofin}

The CEM framework can determine the parameters that lead to a constrained state of static equilibrium $\constrainedequilibriumattributes$ that best satisfies a priori geometric and structural design requirements.
This is accomplished by minimizing an objective function using gradient-based optimization.

\subsubsection{Optimization parameters}

The vector of optimization parameters $\optimizationvariables$ defines the potential solution space of a constrained form-finding problem.
It collects a subset of the design parameters $\designparameters$ (see Section \ref{parameters}).
Design parameters that are not included in $\optimizationvariables$ stay constant throughout the optimization process.

\subsubsection{System solution}
\label{system_solution}

The CEM form-finding algorithm provides an explicit solution $\equilibriumattributes(\optimizationvariables)$ for a given a choice of optimization parameters $\optimizationvariables$.
As per Section \ref{fofin}, this solution contains the missing node positions $\nodeposition$, the internal forces in the trail edges $\trailedgeforce$, the lengths of the deviation edges $\deviationedgelength$, and the reaction forces at the supports $\supportforce$.
The output solution described by $\optimizationvariables$ and $\equilibriumattributes(\optimizationvariables)$ is in static equilibrium.

\subsubsection{Constraints}
\label{constraints}

Vector $\optimizationvariables$ is modified to satisfy nonlinear equality constraints. 
Each constraint $g_i$ is formulated as a function of the optimization parameters and the system solution $\equilibriumattributes(\optimizationvariables)$:

\begin{equation}
\label{eq:constraint}
    \constraint_i(\equilibriumattributes(\optimizationvariables)) = 0
\end{equation}

The constraint functions $g_i$ can be formulated arbitrarily.
However, the complexity of the formulation may affect the solution and the convergence rate of the problem.
We list in Table \ref{constraints_table} the most frequently used geometric- and force-related constraint functions which measure the distance between the current value and a target value for one of the attributes in $\equilibriumattributes(\optimizationvariables)$.
These functions can be freely combined in Equation \ref{opt_problem}.

\begin{table}[!t]
    \centering
    \begin{tabular}{@{}cll@{}}
    \toprule
    
    Type &
    Target &
    Constraint function \\
    
    \midrule
    
    \multirow{3}{*}[-6pt]{Geometry}&
    Node position, $\bar{\nodeposition_i}$&
    $\constraint_1( \equilibriumattributes(\optimizationvariables)) = \norm{\nodeposition_i - \bar{\nodeposition}_i}$\\
    \addlinespace[0.5em]
    
    &
    Edge direction, $\bar{\mathbf{a}}_{i, j}$&
    $\constraint_2(\equilibriumattributes(\optimizationvariables)) = \left|\frac{\nodeposition_j - \nodeposition_i}{\norm{\nodeposition_j - \nodeposition_i}} \cdot \bar{\mathbf{a}}_{i, j}\right| - 1$\\
    \addlinespace[0.5em]
    
    &
    Edge length, $\bar{\edgelength}_{i, j}^{\text{d}}$&
    $\constraint_3(\equilibriumattributes(\optimizationvariables)) = \deviationedgelength_{i, j} - \bar{\edgelength}_{i, j}^{\text{d}}$\\ 
    \addlinespace[0.25em]
    \cmidrule(lr){1-3}
    \addlinespace[0.25em]

    \multirow{3}{*}[-6pt]{Force}&
    Edge force, $\bar{\edgeforce}_{i, j}^{\text{t}}$&
    $\constraint_4(\equilibriumattributes(\optimizationvariables)) = \trailedgeforce_{i, j} - \bar{\edgeforce}_{i, j}^{\text{t}}$\\
    \addlinespace[0.5em]
    
    &
    Edge load path, $\bar{\edgeloadpath}_{i, j}$&
    $\constraint_5(\equilibriumattributes(\optimizationvariables)) = \edgeforce_{i, j}\edgelength_{i, j} - \bar{\edgeloadpath}_{i, j}$\\
    \addlinespace[0.5em]
    
    &
    Reaction force, $\bar{\supportforce}_i$&
    $\constraint_6(\equilibriumattributes(\optimizationvariables)) = \norm{\supportforce_i - \bar{\supportforce}_i}$\\
    \addlinespace[0.5em]
    
    \bottomrule
\end{tabular}
    \caption{Selection of constraint functions supported by the CEM framework.
    From a geometrical vantage point, $\constraint_1$ sets a target position $\bar{\nodeposition}_i$ for node $\node_i$; $\constraint_2$ a target orientation vector $\bar{\mathbf{a}}_{i, j}$ for edge $\edge_{i, j}$; while $\constraint_3$ a target length $\bar{\deviationedgelength}_{i, j}$ for deviation vector $\deviationedge_{i, j}$. 
    Similarly, but from a force perspective, $\constraint_4$ prescribes a desired force magnitude $\bar{\trailedgeforce}_{i, j}$ for trail edge $\trailedge_{i, j}$; $\constraint_5$ a target individual load path $\bar{\edgeloadpath}_{i, j}$ for edge $\edge_{i, j}$; and $\constraint_6$ a target reaction force vector $\bar{\supportforce}_{i}$ at support node $\nodesupport_i$.
    The edge load path $\edgeloadpath_{i, j}$ in $\constraint_5$ corresponds to Maxwell's load path \cite{maxwell_reciprocalfigures_1870}.
    The minimization of this non-negative quantity over all the edges is conducive to a minimum-volume, pin-jointed bar structure \cite{beghini_structuraloptimization_2014, liew_optimisingload_2019a}.
    }
    \label{constraints_table}
\end{table}

\subsubsection{Objective function}
\label{objectivefunction}

Every nonlinear equality constraint $g_i$ is weighted by a penalty factor $w_i$ and aggregated into a single objective function $\penaltyfunction$ that is minimized to solve a constrained form-finding problem.

\begin{equation}
\label{opt_problem}
\mathcal{L}(\optimizationvariables)
=
\frac{1}{2} \sum_i w_i \, g_i( \equilibriumattributes(\optimizationvariables))^2
\end{equation}

\subsubsection{Gradient-based optimization}

Equation \ref{opt_problem} can be efficiently minimized using first-order gradient descent or any other gradient-based optimization algorithm, such as the Limited-Memory Broyden–Fletcher–Goldfarb–Shanno algorithm (L-BFGS) \cite{nocedal_updatingquasinewton_1980}, Sequential Least Squares Programming (SLSQP) \cite{kraft_algorithm733_1994} or Truncated Newton (TNEWTON) \cite{dembo_truncatednewtonalgorithms_1983}.

\subsubsection{Optimization convergence}

The selected optimization algorithm minimizes Equation \ref{opt_problem} over a prescribed number of optimization iterations $\upsilon^{\text{max}}$.
The algorithm converges to an optimal instance of the optimization parameters $\optimizationvariables$ when one of the two conditions given by Equation \ref{eq:optimization_convergence} is fulfilled:

\begin{equation}
  \label{eq:optimization_convergence}
  \begin{aligned}
    \phantom{||\nabla_s}\penaltyoutput\phantom{||}\leq\convergencethreshold\\
    \norm{\gradientvalue}\leq\gradientthreshold
  \end{aligned}
\end{equation}

The objective convergence threshold $\convergencethreshold$ and the gradient convergence threshold $\gradientthreshold$ are two scalars close to zero (for example, $\convergencethreshold=1\times10^{-6}$).
The first condition in Equation \ref{eq:optimization_convergence} indicates that the output value of the objective function $\penaltyoutput$ approaches zero, which implies that the defined constraints $g_i$ satisfy Equation \ref{eq:constraint}.
An instance of $\optimizationvariables$ that fulfills Equation \ref{eq:constraint} does not exist when the supplied constraints contradict each other.
In such case, the optimizer converges to a local minimum of the objective function where the norm of the gradient vanishes, $\norm{\gradientvalue}\leq\gradientthreshold$.
The computation of the gradient is discussed in Section \ref{autograd}.

\section{Extensions to the CEM framework}
\label{extensions}

We extend the CEM framework to overcome the limitations outlined in Section \ref{limitations}: auxiliary trails facilitate the creation of a valid topology diagram $\topology$ and automatic differentiation enables the computation of more reliable and efficient solutions to constrained form-finding problems.
The extended CEM framework is implemented in a standalone design tool.

\subsection{Auxiliary trails}\label{auxiliary_trail}

An auxiliary trail $\auxiliarytrail=\{ \nodeorigin_i, \nodesupport_j \}$ is a short helper trail with an origin node $\nodeorigin_i$ and a support node $\nodesupport_j$ linked by a single trail edge $\trailedge_{i, j}$ of unit length $\trailedgelength_{i,j}=1$.
We automatically attach an auxiliary trail to any node $\node_i$ in a topology diagram $\topology$ that has not been assigned to another trail before the application of the CEM form-finding algorithm (see Section \ref{fofin}).
Such trail-free nodes are characteristic at the intersection between one or more deviation edges and no trail edges.
The attachment operation transforms node $\node_i$ into the origin node $\nodeorigin_i$ of $\auxiliarytrail$.

The extensive use of auxiliary trails enables the explicit construction of the topology diagram $\topology$ of a structure using only deviation edges.
Given an input $\topology$ wherein every bar of a structure is modeled as a deviation edge $\deviationedge_{i,j}$, appending an auxiliary trail to every node $\node_i$ in $\topology$ converts this initially invalid diagram into a diagram $\topology$ that complies with the topological modeling rules of the CEM form-finding algorithm (Section \ref{rules}).
Such a deviation-only modeling strategy circumvents the manual edge labeling process described in Section \ref{topology_diagram} as no distinction has to be made upfront by a designer on whether an edge $\edge_{i, j}$ is a trail edge $\trailedge_{i, j}$ or a deviation edge $\deviationedge_{i, j}$.
We show two examples of structures that are modeled using the deviation-only modeling strategy in Sections \ref{wheel} and \ref{tree}.

However, the topological modeling flexibility enabled by auxiliary trails comes at a computation price.
The attachment of an auxiliary trail creates a local artificial subsystem wherein a specific state of static equilibrium must be computed: the auxiliary trails must not carry any loads in order to capture the originally intended load-carrying behavior of the structure. 
An analogy for the role auxiliary trails play is that they provide additional temporary support to a structure while it is being form-found.
Since we are interested in the "self-standing" version of the structure, an equilibrium state must be found in which the loads these temporary supports carry are zeroed out.

\begin{figure*}[b!]
    \centering
    \hspace{-2.0cm}
    \begin{subfigure}[b]{0.37\textwidth}
        \centering
        \includegraphics[width=\textwidth,trim={0 2mm 0 0},clip]{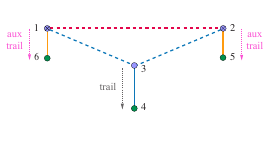}
        \caption{Branching structure A}
        \label{fig:tree_2d_valid_a}
    \end{subfigure}
    \hspace{-1.0cm}
    \begin{subfigure}[b]{0.37\textwidth}
        \centering
        \includegraphics[width=\textwidth,trim={0 2mm 0 0},clip]{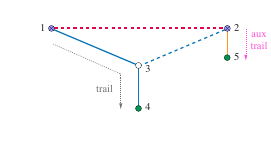}
        \caption{Branching structure B}
        \label{fig:tree_2d_valid_b}
    \end{subfigure}
    \hspace{-1.0cm}
    \begin{subfigure}[b]{0.37\textwidth}
        \centering
        \includegraphics[width=\textwidth,trim={0 2mm 0 0},clip]{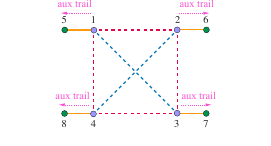}
        \caption{Self-stressed tensegrity}
        \label{fig:self_stressed_valid}
    \end{subfigure}
    \hspace{-2.0cm}
    \caption{Auxiliary trails in a topology diagram $\topology$. To ensure topological validity, one (\ref{fig:tree_2d_valid_b}) and two (\ref{fig:tree_2d_valid_a}) auxiliary trails are appended to the diagram of the branching structure first displayed in Figure \ref{fig:tree_2d_wrong}. Similarly, four auxiliary trails are inserted to that of the self-stressed structure in Figure \ref{fig:self_stressed_valid}.}
    \label{fig:extension_topological_diagrams}
\end{figure*}

Form-finding a structure whose topology diagram $\topology$ contains at least one auxiliary trail hence becomes a constrained form-finding task.
The initially-desired static equilibrium state for a structure is obtained only after solving the optimization problem discussed in Section \ref{constrained_fofin}, where the objective function $\penaltyfunction$ is extended with extra penalty terms we show in Equation \ref{auxiliarytrails_func}.

\begin{equation}\label{auxiliarytrails_func}
  \begin{split}
  \frac{1}{2} \sum_a \weight_a \medspace (\edgeforce_{i, j}^{\text{t}} - \bar{\edgeforce}_{i, j}^{\text{t}})_{a}^2
  \end{split}
\end{equation}

The purpose of the additional penalty terms is to minimize the difference between the target force $\bar{\edgeforce}_{i, j}^{\text{t}}$ and the actual force $\edgeforce_{i, j}^{\text{t}}$ in the trail edge $\trailedge_{i, j}$ of each auxiliary trail $\auxiliarytrail$ in $\topology$.
These terms are equivalent to constraint function $\constraint_4$ in Table \ref{constraints_table}.
By extension, when the force in the trail edge of an auxiliary trail is zero $\edgeforce_{i, j}^{\text{t}} = 0$, the reaction forces incident to its corresponding support node also vanish, $\supportforce=\mathbf{0}$.

Figure \ref{fig:extension_topological_diagrams} shows three topology diagrams $\topology$ that use auxiliary trails to remedy the modeling challenges posed by the diagrams in Figure \ref{fig:topological_diagrams}.
Figure \ref{fig:tree_2d_valid_a} depicts a diagram $\topology$ where auxiliary trails $\auxiliarytrail_2=\{1, 6\}$ and $\auxiliarytrail_3=\{2, 5\}$ are appended to nodes 1 and 2.
Meanwhile, in Figure \ref{fig:tree_2d_valid_b}, trail $\trail_2 = \{2, 3, 4\}$ is deleted, thus leaving node $\node_2$ trail unassigned.
An auxiliary trail $\auxiliarytrail_2=\{2, 5\}$ is attached to $\node_2$ to make $\topology$ valid again.
As portrayed by Figure \ref{fig:self_stressed_valid}, an auxiliary trail is annexed to each of the four nodes of the tensegrity structure in Figure \ref{fig:self_stressed_wrong} to correct its initial topological invalidity.

\subsection{Automatic and exact computation of the gradient}
\label{autograd}

The gradient required to determine a minimum of Equation \ref{opt_problem} results from the first derivative of $\mathcal{L}$ with respect to the optimization parameters:

\begin{equation}
\label{gradient_func}
    \gradient
    =
    \sum_i w_i \, g_i \cdot \nabla_{\optimizationvariables} \, g_i
\end{equation}

The computation of $\gradient$ requires the calculation of the derivatives of the individual constraint functions $\nabla_{\optimizationvariables} \, g_i$ and the derivative of the system solution $\equilibriumattributes(\optimizationvariables)$:

\begin{equation}
    \nabla_{\optimizationvariables} \, g_i
    =
    \frac{\partial{g_i}}{\partial{\equilibriumattributes}}
    \frac{\partial{\equilibriumattributes}}{\partial{\mathbf{s}}}
\end{equation}

While Equations \ref{node_xyz}-\ref{equilibrium_energy} are compact algebraic manipulations for which derivatives can be found analytically, manually applying the chain rule through the control flow structure of the CEM form-finding algorithm (see Algorithm \ref{fofin_algo}) to calculate the partial derivative of the system solution with respect to the optimization parameters has been a complex task \cite{ohlbrock_computeraidedapproach_2020}.
Instead of circumventing its sequential and iterative characteristics, we exploit the implementation of the CEM form-finding algorithm we develop in Section \ref{tool} by using AD in reverse mode to obtain a version of $\gradientvalue$ that is exact up to floating-point precision.
AD ingests the function $\penaltyfunction$, and generates another function that calculates the associated gradient $\gradientvalue$.
The key insight is that $\gradientvalue$ is automatically generated by a computer program.

We stress that AD provides a numerical value of $\gradientvalue$ evaluated at a specific instance of $\optimizationvariables$ instead of generating an analytical expression for it.
Nevertheless, the AD output is adequate for our purposes since we use the value of $\gradientvalue$ to minimize Equation \ref{opt_problem}, regardless of what the underlying analytical expression might be.
We refer the reader to \cite{corliss_automaticdifferentiation_2013, baydin_automaticdifferentiation_2018} for a detailed theoretical treatment on how reverse-mode AD evaluates derivatives of algorithmically expressed functions.
To illustrate how AD operates specifically through the calculations of the CEM framework, we supplement the discussion with a toy constrained form-finding example.

\subsubsection{Example}
\label{ad_example}

\begin{figure}[t!]
    \centering
    \begin{subfigure}[b]{\columnwidth}
        \centering
        \includegraphics[width=\textwidth]{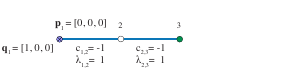}
        \caption{Topology diagram $\topology$, edge internal force states $\internalforcestate_{1,2},\internalforcestate_{2,3}$; and input design parameters $\designparameters$ (the position $\nodeposition_1$ of origin node $\nodeorigin_1$; the edge lengths $\edgelength_{1,2},\edgelength_{2,3}$; and the applied load $\nodeload_1$).}
        \label{fig:strut_topology}
    \end{subfigure}
    
    \bigskip
    
    \begin{subfigure}[b]{\columnwidth}
        \centering
        \includegraphics[width=\textwidth]{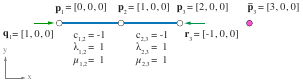}
        \caption{Form diagram $\form$ and the equilibrium state $\equilibriumattributes$ generated with the CEM form-finding algorithm, which consists of the node positions $\nodeposition_2,\nodeposition_3$; the edge absolute internal forces $\edgeforce_{1,2}, \edgeforce_{2,3}$; and the support reaction force $\supportforce_3$. To solve the constrained form-finding problem, position $\nodeposition_3$ should reach the target position $\bar{\nodeposition}_3$.
        }
        \label{fig:strut_form}
    \end{subfigure}
    
    \bigskip
    
    \begin{subfigure}[b]{\columnwidth}
        \centering
        \includegraphics[width=\textwidth]{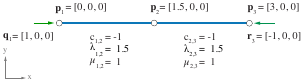}
        \caption{Constrained form diagram $\constrainedform$ and the equilibrium state $\constrainedequilibriumattributes$ post-optimization. The node position $\nodeposition_3$ moves to $\bar{\nodeposition}_3$ once the gradient-based optimizer modifies the edge lengths such that $\edgelength_{1,2}=\edgelength_{2,3}=1.5$.
        }
        \label{fig:strut_form_constrained}
    \end{subfigure}
    \caption{Constrained form-finding of a two-segment strut.}
\end{figure}

Consider a two-segment strut subjected to an horizontal compressive force $\nodeload_1$.
Figure \ref{fig:strut_topology} shows the topology diagram $\topology$, the internal force states $\internalforcestates$ and the design parameters $\designparameters$.
The compression-only diagram $\form$ corresponding to the state of static equilibrium $\equilibriumattributes$ output by the CEM form-finding algorithm is given in Figure \ref{fig:strut_form}.
In this example, we impose a geometric restriction on the position of node $\node_3$: it should land at $\bar{\nodeposition}_3=[3,0,0]$.
However, the resulting position $\nodeposition_3=[2,0,0]$ is away from the target.

We are deliberately unsure of what combination of trail edge lengths $\trailedgelength_{i,j}$ would result in a constrained state of static equilibrium $\constrainedequilibriumattributes$ that matches $\bar{\nodeposition}_3$.
To solve this constrained form-finding problem, we set the two trail edge lengths as optimization variables such that $\optimizationvariables=[\edgelength_{1,2},\edgelength_{2, 3}]=[1,1]$.
The superscript ${\text{t}}$ in $\trailedgelength_{i, j}$ is dropped in this example for legibility.
The only constraint function used is $\constraint_1$ from Table \ref{constraints_table} and the penalty factor $\weight_1$ is set to $\weight_1=1$.

The value of the gradient that reverse-mode AD computes is $\gradientvalue=[-1,-1]$.
As expected, the negative magnitude of the partial derivatives in the gradient indicates that adjusting $\optimizationvariables$ in the opposite direction of the gradient would elongate both trail edges for $\nodeposition_3$ to move closer to the target position $\bar{\nodeposition}_3$ after the next optimization step.
Reverse-mode AD arrives at $\gradientvalue=[-1,-1]$ after processing one forward and one reverse computation trace, one after the other, as we show in Table \ref{table:grad_strut_example}.

\begin{figure}[b!]
    \centering
    \begin{tikzpicture}

\tikzstyle{io} = [circle, minimum size=6mm]
\tikzstyle{eq} = [circle, minimum size=6mm, fill=orange!30]
\tikzstyle{constraint} = [circle, minimum size=6mm, fill=teal!30] 
\tikzstyle{objective} = [circle, minimum size=6mm, fill=pink!60] 
\tikzstyle{extra} = [circle, minimum size=6mm, fill=black!0] 
\tikzstyle{arrow} = [-{Latex[length=1.mm]}, shorten >=0.5pt, rounded corners]
\tikzstyle{arrowback} = [densely dotted, {Latex[length=1.mm]}-, shorten >=0.5pt, rounded corners]

\node (v1) [eq] at (0, 0) {$v_1$};
\node (v3) [eq] at (0, -1.2) {$v_3$};

\node (v2) [eq, right=0.4 of v1] {$v_2$};
\node (v4) [eq, right=0.4 of v3] {$v_4$};
\node (v5) [constraint, right=0.4 of v4] {$v_5$};
\node (v6) [constraint, right=0.4 of v5] {$v_6$};
\node (v7) [constraint, right=0.4 of v6] {$v_7$};
\node (v8) [objective, below=0.4 of v7] {$v_8$};
\node (v9) [objective, right=0.4 of v8] {$v_9$};

\node (input1) [io, left=0.4 of v1] {$\edgelength_{1, 2}$};
\node (input2) [io, left=0.4 of v3] {$\edgelength_{2, 3}$};
\node (output) [io, right=0.4 of v9] {$\penaltyoutput$};

\node (r1) [extra, above=0.4 of v1] {$\noderesidual_1$};
\node (p1) [extra, above=0.4 of v2] {$\nodeposition_1$};
\node (r2) [extra, below=0.4 of v3] {$\noderesidual_2$};
\node (y3) [extra, below=0.4 of v5] {$\bar{\nodeposition}_3$};

\draw[arrow] ($(input1.east)+(0,0.1)$) -- ($(v1.west)+(0,0.1)$);
\draw[arrow] ($(input2.east)+(0,0.1)$) -- ($(v3.west)+(0,0.1)$);

\draw[arrow] ($(v1.east)+(0,0.1)$) -- ($(v2.west)+(0,0.1)$);
\draw[arrow] ($(v3.east)+(0,0.1)$) -- ($(v4.west)+(0,0.1)$);
\draw[arrow] ($(v2.south)+(0.1,0)$) -- ($(v4.north)+(0.1,0)$);
\draw[arrow] ($(v4.east)+(0,0.1)$) -- ($(v5.west)+(0,0.1)$);
\draw[arrow] ($(v5.east)+(0,0.1)$) -- ($(v6.west)+(0,0.1)$);
\draw[arrow] ($(v6.east)+(0,0.1)$) -- ($(v7.west)+(0,0.1)$);
\draw[arrow] ($(v7.south)+(0.1, 0)$) -- ($(v8.north)+(0.1,0)$);

\draw[arrow] ($(v8.east)+(0,0.1)$) -- ($(v9.west)+(0,0.1)$);
\draw[arrow] ($(v9.east)+(0,0.1)$) -- ($(output.west)+(0,0.1)$);

\draw[arrow] (r1) -- (v1);
\draw[arrow] (p1) -- (v2);
\draw[arrow] (r2) -- (v3);
\draw[arrow] (y3) -- (v5);

\draw[arrowback] ($(input1.east)-(0,0.1)$) -- ($(v1.west)-(0,0.1)$);
\draw[arrowback] ($(input2.east)-(0,0.1)$) -- ($(v3.west)-(0,0.1)$);

\draw[arrowback] ($(v1.east)-(0,0.1)$) -- ($(v2.west)-(0,0.1)$);
\draw[arrowback] ($(v3.east)-(0,0.1)$) -- ($(v4.west)-(0,0.1)$);
\draw[arrowback] ($(v2.south)-(0.1,0)$) -- ($(v4.north)-(0.1,0)$);
\draw[arrowback] ($(v4.east)-(0,0.1)$) -- ($(v5.west)-(0,0.1)$);
\draw[arrowback] ($(v5.east)-(0,0.1)$) -- ($(v6.west)-(0,0.1)$);
\draw[arrowback] ($(v6.east)-(0,0.1)$) -- ($(v7.west)-(0,0.1)$);
\draw[arrowback] ($(v7.south)-(0.1, 0)$) -- ($(v8.north)-(0.1,0)$);

\draw[arrowback] ($(v8.east)-(0,0.1)$) -- ($(v9.west)-(0,0.1)$);
\draw[arrowback] ($(v9.east)-(0,0.1)$) -- ($(output.west)-(0,0.1)$);

\end{tikzpicture}
    \caption{Computation graph of the two-segment strut example described in Section \ref{ad_example}.
    The graph traces the operations involved in the evaluation the objective function $\penaltyoutput$ at optimization parameters $\optimizationvariables=[\edgelength_{1, 2}, \edgelength_{2, 3}]$ during the forward pass (solid arrows).
    The nodes $v_i$ store the output of each of the intermediary operations that modify $\optimizationvariables$ on their way to $\penaltyoutput$.
    Nodes $v_1$ to $v_4$ correspond to operations that occur within the CEM form-finding algorithm (see Algorithm \ref{fofin_algo}), while nodes $v_5$ to $v_7$ to those executed in the evaluation of the constraint function $\constraint_1$.
    Nodes $v_8$ and $v_9$ evaluate Equation \ref{opt_problem}.
    To evaluate the gradient $\gradientvalue$, reverse-mode AD propagates the partial derivatives of each node as per Equation \ref{eq:adjoint} in the opposite direction of the forward pass (backpropagation, dotted arrows).
    We unpack the operations of both the evaluation trace and the derivatives trace in Table \ref{table:grad_strut_example}.}
\label{fig:strut_computation_graph}
\end{figure}
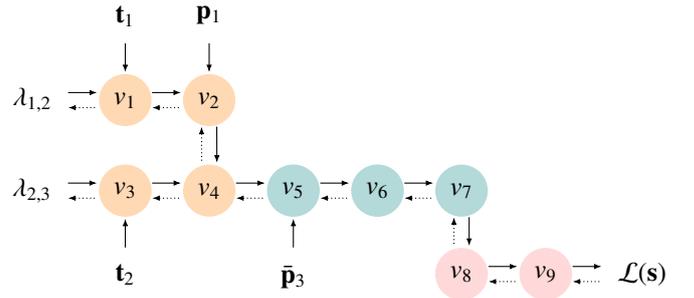

First, AD builds a forward evaluation trace (also called a Wenger list \cite{wengert_simpleautomatic_1964}) once it calculates the output value of the objective function $\penaltyoutput$.
During the assembly of the evaluation trace, AD keeps track of the sequence of operations that interact with the entries in $\optimizationvariables$ in their journey towards $\penaltyoutput$.
The evaluation trace is broken next into a sequence of elementary mathematical operations, such as sums, divisions and multiplications, whose outputs are stored in intermediary variables $v_i$.
Dependency relations between the variables $v_i$ are finally represented as nodes and edges in a computation graph \cite{bauer_computationalgraphs_1974} which we show for this example in Figure \ref{fig:strut_computation_graph}.
In this example, from the 42 steps that Algorithm \ref{fofin_algo} comprises, only one modifies $\optimizationvariables$, which corresponds to Equation \ref{node_xyz}.
We highlight the capability of AD to identify and register this automatically despite the loops and conditional statements in Algorithm \ref{fofin_algo}.

Once the evaluation trace is complete, reverse-mode AD backpropagates derivatives on the nodes of the computation graph that displayed in Figure \ref{fig:strut_computation_graph}.
We present the derivatives trace of our example on the right-hand side of Table \ref{table:grad_strut_example}.
This reverse derivatives trace starts off at the value of the objective function $\penaltyoutput$ and ends once the nodes in the graph that correspond to the optimization parameters $\optimizationvariables$ are reached.
Unlike forward-mode AD, only one pass over the entire computation graph suffices to compute $\gradientvalue$ \cite{griewank_evaluatingderivatives_2008}.

As AD walks in reverse over the graph, it calculates the partial derivative of $\penaltyoutput$ with respect to every intermediate node using the chain rule.
This partial derivative $\bar{v}_i$, called an \emph{adjoint}, quantifies the sensitivity of the output $\penaltyoutput$ to changes in the value of an intermediary variable $v_i$ as expressed in Equation \ref{eq:adjoint}.

\begin{equation}
\label{eq:adjoint}
  \begin{split}
    \bar{v}_i
    =
    \frac{\partial{\penaltyoutput}}{\partial{v_i}}
    =
    \sum_{j} \bar{v}_j \frac{\partial{v_j}}{\partial{v_i}}
  \end{split}
\end{equation}

The calculation of $\bar{v}_i$ is carried out by looking at each of the $j$ immediate children nodes of the variable $v_i$ in the computation graph \cite{bauer_computationalgraphs_1974, oktay_randomizedautomatic_2020}.
In this example, the value of the gradient finally results from the adjoints of the optimization parameters $\gradientvalue=[\bar{\edgelength}_{1, 2}, \bar{\edgelength}_{2, 3}]$ where their partial derivatives are scaled by the adjoints $\bar{v}_1$ and $\bar{v}_3$, such that $\bar{\edgelength}_{1, 2}=\bar{v}_1\:\partial{v_1}/\partial{\edgelength_{1, 2}}$ and $\bar{\edgelength}_{2, 3}=\bar{v}_3\:{\partial{v_3}}/\partial{\edgelength_{2, 3}}$.

\begin{table*}[ht!]
    \centering
    \begin{tabular}{@{}llllllllll@{}}
    \toprule
    \multicolumn{5}{l}{\textbf{Evaluation trace (Forward pass)}} &
    \multicolumn{4}{l}{\textbf{Derivatives trace (Backpropagation)}}\\
    
    \midrule
    
    \tikzmark{a} &
    $ \edgelength_{1, 2}$  & 
    &
    &
    $= 1$ &
    \tikzmark{c} &
    $\bar{\edgelength}_{1, 2}$ &
    $= \bar{v}_1\frac{\partial{v_1}}{\partial{\edgelength_{1, 2}}}$  &
    $= \bar{v}_1 ^\top \internalforcestate_{1, 2} \frac{\noderesidual_{1}}{\norm{\noderesidual_{1}}}$ &
    $= -1$ \\
    
    &
    $ \edgelength_{2, 3}$  & 
    &
    &
    $= 1$ &
    &
    $\bar{\edgelength}_{2, 3}$ &
    $= \bar{v}_3\frac{\partial{v_3}}{\partial{\edgelength_{2, 3}}}$  &
    $= \bar{v}_3 ^\top \internalforcestate_{2, 3} \frac{\noderesidual_{2}}{\norm{\noderesidual_{2}}}$ &
    $= -1$ \\
    
    \cmidrule(lr){2-5}\cmidrule(lr){7-10}
    
    &
    $v_1$ & 
    $= \edgelength_{1, 2} \times \internalforcestate_{1, 2} \frac{\noderesidual_{1}}{\norm{\noderesidual_{1}}}$ & 
    $= 1 \times \frac{[1, 0, 0]}{1}$ & 
    $= [\phantom{-}1, 0, 0]$ &
    &
    $\bar{v}_1$ &
    $= \bar{v}_2 \frac{\partial{v_2}}{\partial{v_1}}$  &
    $= \bar{v}_2 \times 1$ &
    $= [-1, 0, 0]$ \\
    
    &
    $v_2$ & 
    $= v_1 + \nodeposition_1$ &
    $= [1, 0, 0] + [0, 0, 0]$ &
    $= [\phantom{-}1, 0 ,0]$ &
    &
    $\bar{v}_2$ &
    $= \bar{v}_4 \frac{\partial{v_4}}{\partial{v_2}}$  &
    $= \bar{v}_4 \times 1$ &
    $= [-1, 0, 0]$ \\
    
    &
    $v_3$ & 
    $= \edgelength_{2, 3} \times \internalforcestate_{2, 3} \frac{\noderesidual_{2}}{\norm{\noderesidual_{2}}}$ &
    $= 1 \times \frac{[1, 0, 0]}{1}$ & 
    $= [\phantom{-}1, 0, 0]$ &
    &
    $\bar{v}_3$ &
    $= \bar{v}_4 \frac{\partial{v_4}}{\partial{v_3}}$  &
    $= \bar{v}_4 \times 1$ &
    $= [-1, 0, 0]$ \\
    
    &
    $v_4$ & 
    $= v_3 + v_2$ &
    $= [1, 0, 0] + [1, 0, 0]$ &
    $= [\phantom{-}1, 0, 0]$ &
    &
    $\bar{v}_4$ &
    $= \bar{v}_5 \frac{\partial{v_5}}{\partial{v_4}}$  &
    $= \bar{v}_5 \times 1$ &
    $= [-1, 0, 0]$ \\
    
    &
    $v_5$ &
    $= v_4 - \bar{\nodeposition}_3$ & 
    $= [2, 0, 0] - [3, 0, 0]$ & 
    $= [-1, 0, 0]$ &
    &
    $\bar{v}_5$ & 
    $= \bar{v}_6 \frac{\partial{v_6}}{\partial{v_5}}$  &
    $= \bar{v}_6 \times 2 \times v_5$ &
    $= [-1, 0, 0]$ \\
    
    &
    $v_6$ & 
    $= v_5 ^\top \cdot v_5$ &
    $= [-1, 0, 0]^\top  \cdot[-1, 0, 0]$ & 
    $= 1$ &
    &
    $\bar{v}_6$ & 
    $= \bar{v}_7 \frac{\partial{v_7}}{\partial{v_6}}$ &
    $= \bar{v}_7 \times 0.5 \times \sqrt{v_6}$ &
    $= 0.5$ \\
    
    &
    $v_7$ & 
    $=\sqrt{v_6}$ &
    $=\sqrt{1}$ &
    $= 1$ &
    &
    $\bar{v}_7$ & 
    $= \bar{v}_8 \frac{\partial{v_8}}{\partial{v_7}}$ &
    $= \bar{v}_8 \times 2 \times v_7$ &
    $= 1$ \\
    
    &
    $v_8$ & 
    $=v_7 \times v_7$ &
    $=1 \times 1$ &
    $= 1$ &
    &
    $\bar{v}_8$ & 
    $= \bar{v}_9 \frac{\partial{v_9}}{\partial{v_8}}$ &
    $= \bar{v}_9 \times 0.5$ &
    $= 0.5$ \\
    
    &
    $v_9$ & 
    $=v_8 \times 0.5$ &
    $=1 \times 0.5$ &
    $= 0.5$ &
    &
    $\bar{v}_9$ & 
    $= \bar{\penaltyfunction}(\optimizationvariables) \frac{\partial{\penaltyoutput}}{\partial{v_9}}$ &
    $= \bar{\penaltyfunction}(\optimizationvariables) \times 1$ &
    $= 1$ \\
    
    \cmidrule(lr){2-5}\cmidrule(lr){7-10}
    
    \tikzmark{b} &
    $\penaltyoutput$ & 
    &
    &
    $= 0.5$ &
    \tikzmark{d} &
    $\bar{\penaltyfunction}(\optimizationvariables)$ &
    &
    & 
    $= 1$\\
    
    \bottomrule
\end{tabular}
\tikz[remember picture,overlay] \draw[->, -{Latex[length=2.5mm]}, thick] (a.center -| b.center) -- (b.center);
\tikz[remember picture,overlay] \draw[->, -{Latex[length=2.5mm]}, thick, dotted] (d.center -| c.center) -- (c.center);
    \caption{
    Automatic differentiation (AD) applied to the CEM form-finding algorithm Equation \ref{opt_problem} to solve the constrained form-finding problem depicted in Figure \ref{fig:strut_form}.
    To evaluate the gradient $\gradientvalue$ of Equation \ref{opt_problem} in Section \ref{ad_example}, reverse-mode AD operates on two computation traces: firstly, one evaluation trace (left-hand side) and secondly, one derivatives trace (right-hand side).
    To construct the former, AD evaluates Equation \ref{opt_problem} and Algorithm \ref{fofin_algo}, and keeps of all the elementary operations that modify the optimization parameters $\optimizationvariables=[\edgelength_{1, 2}, \edgelength_{2, 3}]$ (in this case, the length of the edges $\edge_{1, 2}$ and $\edge_{2, 3}$), and the sequence in which they alter them.
    The output of each basic operation is stored in intermediary variables $v_i$ which finally become interconnected nodes in the computation graph we show in Figure \ref{fig:strut_computation_graph}.
    AD calculates partial derivatives of each of the nodes with respect to $\penaltyoutput$ (i.e. the \emph{adjoints}, $\bar{v}_i$) walking in reverse over the edges of the graph and applying the chain rules.
    The walk starts from the last operation tracked in the evaluation trace and ends when $\optimizationvariables$ is reached.
    The adjoints of the optimization parameters are finally the entries in the gradient, $\gradientvalue=[-1, -1]$.
    After \cite{baydin_automaticdifferentiation_2018}.}
    \label{table:grad_strut_example}
\end{table*}


\subsection{Design tool}\label{tool}

\usemintedstyle{emacs}
\begin{figure}[!b]
    \inputminted[
    xleftmargin=14pt,
    numbersep=7pt,
    frame=lines,
    framesep=2mm,
    fontsize=\footnotesize,
    linenos
    ]{python}{code/edge_chain.py}
\caption{Python code that models the compression chain shown in Figures \ref{fig:strut_topology} and \ref{fig:strut_form} with the version of \texttt{compas\_cem} at the time of writing \cite{compas-cem}.}
\label{fig:compascemcode}
\end{figure}

With the goal of making our work usable and reproducible, the CEM framework and the extensions presented hitherto are consolidated in a standalone, open-source design tool called \texttt{compas\_cem} \cite{compas-cem}.
The tool is written in Python \cite{python} and is integrated into the COMPAS framework, a computational ecosystem for collaboration and research in architecture, engineering, fabrication, and construction \cite{compas-dev}.
As a COMPAS extension, \texttt{compas\_cem} can interface seamlessly with other packages in the COMPAS framework to perform additional tasks on the structures generated with this tool \cite{compas-fea, compas-vol, compas-fab}.

A first CEM toolkit was presented in \cite{cem_tool_2021} as a plugin bound to the Windows version of Grasshopper \cite{grasshopper}.
In contrast, \texttt{compas\_cem} runs independently from 3D modeling software, and it makes it possible to solve constrained form-finding problems on tension-compression structures readily from the command line interface of any of three major computer operating systems: Windows, MacOS and Linux.
Furthermore, COMPAS provides our tool with the necessary interfaces to be invoked directly inside Blender \cite{blender}, Rhino for Windows and Rhino for MacOS \cite{rhino3d}, and Grasshopper \cite{grasshopper}.
We illustrate this possibility in the structural design application we discuss in Section \ref{casestudy}.

Currently, \texttt{compas\_cem} uses the implementation of the optimization algorithms in the \texttt{NLopt} library \cite{johnson_nloptnonlinearoptimization_2021} to minimize Equation \ref{opt_problem}, and delegates the evaluation of the gradient $\gradientvalue$ shown in Equation \ref{gradient_func} with AD to \texttt{autograd} \cite{maclaurin_autograd_2015}.
This choice of dependencies is not restrictive.
In the future, we envision integrating other Python optimization and automatic differentiation libraries such as \texttt{scipy} \cite{scipy} and \texttt{hyperjet} \cite{hyperjet}.

The codebase of the design tool we propose follows a modular and object-oriented structure. 
For a granular overview of the objects and functions that it comprises, we refer the interested reader to the latest version of the \texttt{compas\_cem} manual available online \cite{compas-cem}.
We offer instead a walk-trough over a minimal working example of \texttt{compas\_cem} written in 59 lines of Python code and discuss how it relates to the theoretical concepts developed in Sections \ref{cemf} and \ref{extensions}.
This example, shown in Figure \ref{fig:compascemcode}, generates the structure we presented in Section \ref{autograd}. 

The required \texttt{compas\_cem} imports occur in lines 1-11. 
A \texttt{TopologyDiagram()} object, instantiated in line 16 is a child class of the \texttt{Network()} relational datastructure from COMPAS.
A \texttt{Network()} is a graph that facilitates the storage of attributes on its vertices and edges as Python dictionaries. 
Therefore, the inputs to the CEM form-finding algorithm, the topology diagram $\topology$ and the design parameters $\designparameters$, are stored in the same \texttt{TopologyDiagram()} object.
In lines 25-26, we set the internal force state of the trail edges $\internalforcestate_{i,j}$ as the sign of the length that parametrizes them (see Section \ref{topology_diagram}).
The negative length indicates that the edges are under compression, $\internalforcestate_{1,2}=\internalforcestate_{2,3}=-1$ .
A positive length would conversely assign them a tension state.
After defining nodes, edges, supports and loads in lines 18-32, the composition of $\topology$ can be formally expressed as
$\nodes=\{1, 2, 3\}$, $\edges=\{(1, 2), (2, 3)\}$, $\supports=\{3\}$.

A graph traversal algorithm that automatically searches for trails $\trails$ is invoked on line 35.
This algorithm takes the support node $\nodesupport_3$ as the search starting point and moves recursively over the next trail edge $\edge_{i, j}$ connected until no more trail edges are found.
If the boolean argument \texttt{auxiliary\_trails} were set to \texttt{True}, the trail-search algorithm would attach an auxiliary trail $\auxiliarytrail$ to any node that was not automatically assigned to a trail by the graph traversal, as exposed in Section \ref{auxiliary_trail}.
Only one trail is expected in this example, $\trail_1=\{1, 2, 3\}$.
In line 39, the CEM form-finding algorithm is invoked to output a form diagram $\form$, conditioned on $\iterationsmax=1$ and $\energythreshold=1\times10^{-5}$.
As with \texttt{TopologyDiagram()}, $\form$ is a subclass of \texttt{Network()} that self-contains the attributes that describe the generated state of static equilibrium $\equilibriumattributes$.

The generation of a constrained form diagram $\constrainedform$ is spread over lines 43-54. 
Equation \ref{opt_problem} is minimized with the SLSQP optimization algorithm \cite{kraft_algorithm733_1994} from \texttt{NLopt} \cite{johnson_nloptnonlinearoptimization_2021} using an optimization convergence threshold of $\convergencethreshold=1\times10^{-6}$.
In line 46, we specify that the desired position for node $\node_3$ is $\bar{\nodeposition}_3 = [3, 0, 0]$, and in lines 50-51, we define the optimization parameters $\optimizationvariables$ as he length of the trail edges created in lines 25-26 such that $\optimizationvariables = [\trailedgelength_{1, 2}, \trailedgelength_{2, 3}]$.
The constrained form-finding problem is solved in line 54.
Internally, \texttt{autograd} \cite{maclaurin_autograd_2015} evaluates the gradient, as required by the optimization process.
We plot the resulting instance of $\constrainedform$ in line 59 of the code and show it in Figure \ref{fig:strut_form_constrained}.

\section{Numerical validation}
\label{experiments}

The intent of this section is to quantitatively benchmark the extensions we make to the CEM framework.
We study three structures that leverage auxiliary trails to be topologically compatible with the CEM form-finding algorithm: a self-stressed tensegrity wheel, a tree canopy and a bridge curved on plan.

We assume that all the structures are pin-jointed and only bear axial forces.
We model the first two structures using only deviation edges to illustrate how, in an extreme case, inserting an auxiliary trail to every node in $\topology$ can relieve designers from the trail-deviation edge labeling process (see Section \ref{topology_diagram}).
The bridge structure follows a more conventional topological modeling approach and only appends auxiliary trails at the tip of the cantilevering hangers.

The primary goal of all the constrained form-finding experiments in this section is to minimize the forces in the auxiliary trails by setting the target edge force $\hat{\edgeforce_i}$ to zero.
We impose additional geometric constraints to the bridge to test the auxiliary trails extension we propose in combination with more constraint types.
The penalty factors for all the constraints are equal to one, $w=1$, and the distance threshold for iterative equilibrium in the CEM form-finding algorithm is $\energythreshold=1\times10^{-6}$ (See Section \ref{iterative}).

We solve each constrained form-finding experiment using automatic differentiation (AD) and finite differences (FD) with different step sizes $h$, following the implementation of the baseline version of the CEM framework \cite{ohlbrock_computeraidedapproach_2020,ohlbrock_combinatorialequilibrium_2020, cem_tool_2021}.
We fix the optimization convergence thresholds to $\convergencethreshold=1\times10^{-6}$ and $\gradientthreshold=1\times10^{-8}$ in all experiments, and compare the impact of the two differentiation schemes by looking at the total convergence runtime (i.e. the elapsed time in seconds it takes an optimization algorithm to converge), the number of optimization parameters, and the output value of the objective function after convergence to a constrained equilibrium state $\constrainedequilibriumattributes$.
We run every experiment ten times and report the resulting mean values per experiment.

For reference, we execute the work we present in this section for both AD and FD in a collection of jupyter notebooks \cite{kluyver_jupyternotebooks_2016} using \texttt{compas\_cem} on MacOS, on a quad-core Intel CPU clocked at 2.9 GHz.
We make these notebooks available as supplementary data in \cite{cemad-cad}.

\subsection{Auxiliary trails in 2D}
\label{wheel}

\begin{figure}[t!]
    \captionsetup[subfigure]{justification=centering}
    \centering
    \begin{subfigure}[b]{\columnwidth}
    \centering
    \begin{subfigure}[b]{0.45\columnwidth}
        \centering
        \includegraphics[width=0.95\textwidth]{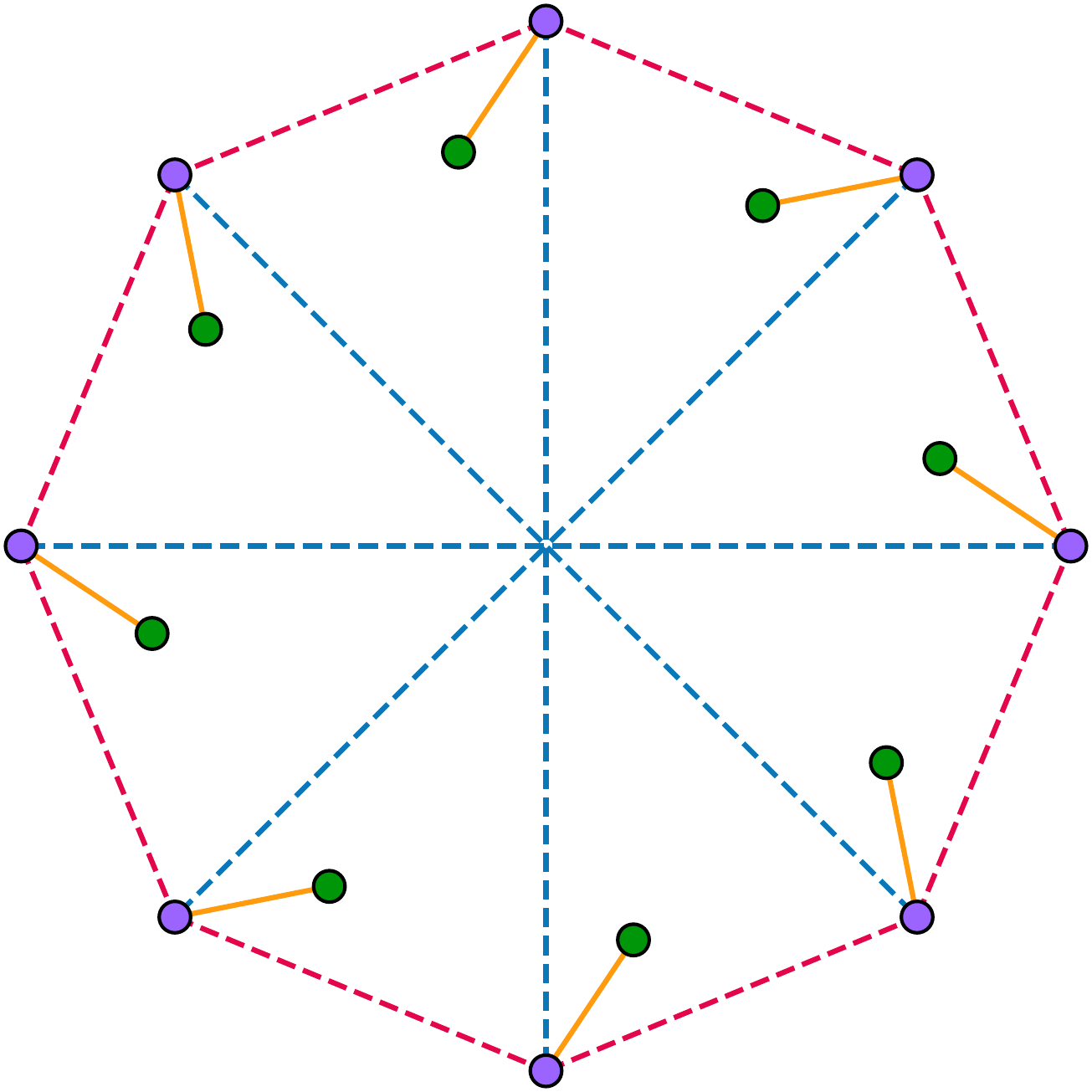}
        \vspace{2mm}
    \end{subfigure}
    \begin{subfigure}[b]{0.45\columnwidth}
        \centering
        \includegraphics[width=\textwidth]{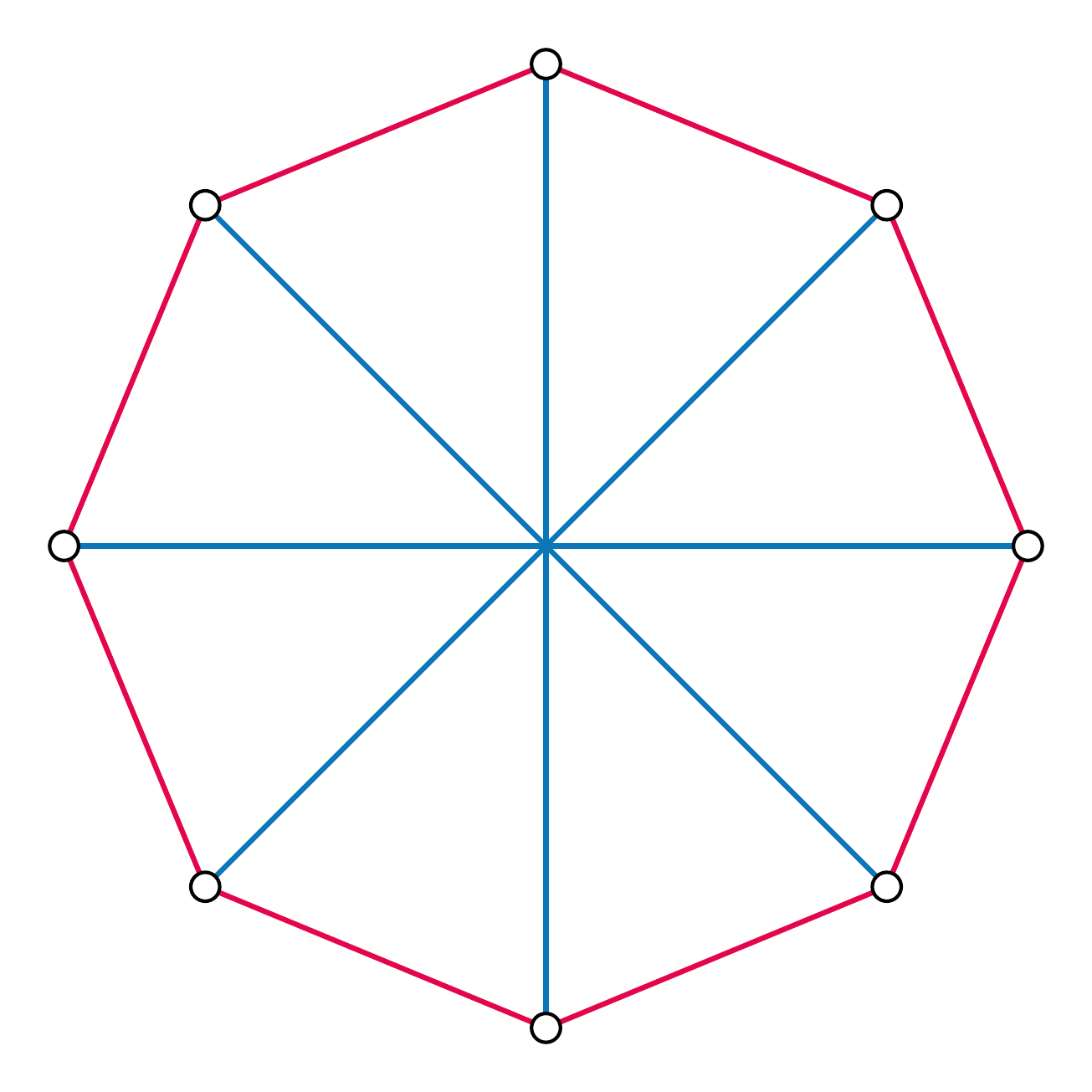}
    \end{subfigure}
    \caption{Topology $\topology$ (left) and constrained form diagram $\constrainedform$ (right) of a wheel.\\12 parameters (8 sides)\\AD: 0.04 sec / FD: 0.09 sec}
    \end{subfigure}

    \begin{subfigure}[b]{\columnwidth}
    \centering
    \vspace{3mm}
    \begin{subfigure}[b]{0.45\columnwidth}
        \centering
        \includegraphics[width=\textwidth]{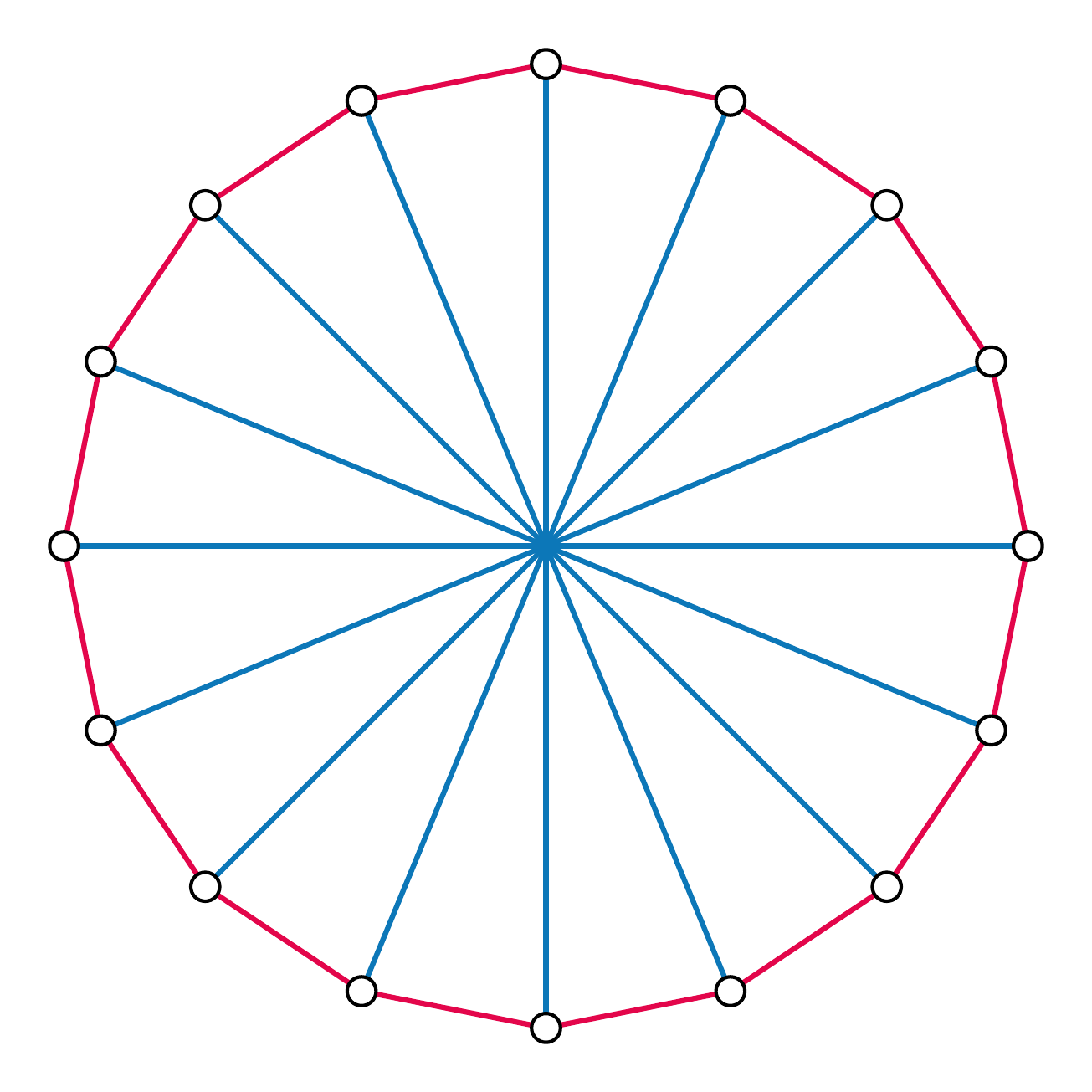}
        \caption*{24 parameters (16 sides)\\AD: 0.12 sec / FD: 0.30 sec}
    \end{subfigure}
    \begin{subfigure}[b]{0.45\columnwidth}
        \centering
        \includegraphics[width=\textwidth]{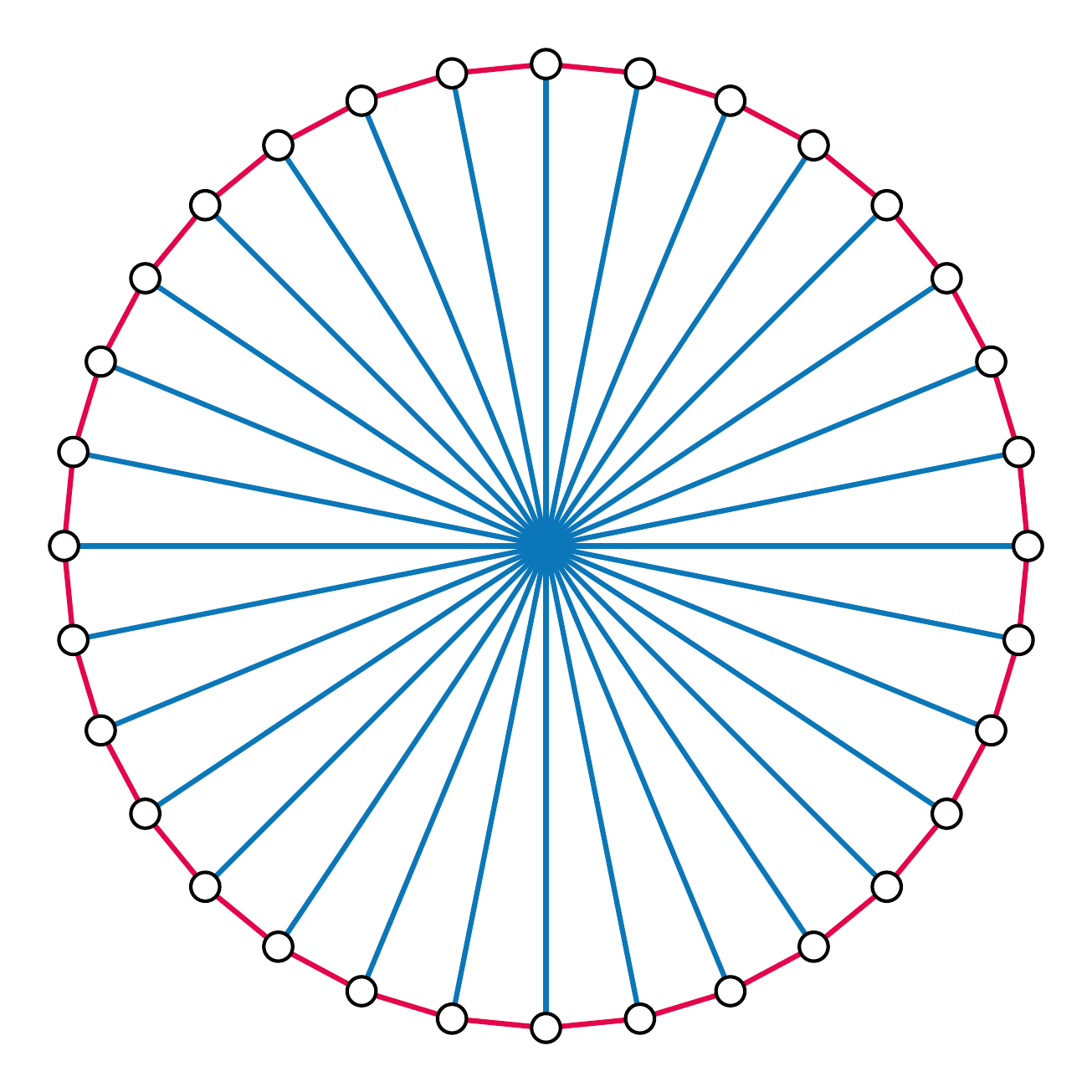}
        \caption*{48 parameters (32 sides)\\AD: 0.17 sec / FD: 1.08 sec}
    \end{subfigure}
    \caption{Constrained form diagrams $\constrainedform$}
    \end{subfigure}
    \caption{Planar tensegrity wheels form-found using the extended CEM framework. The wheels are self-stressed. Auxiliary trails are not drawn in the diagrams $\constrainedform$ as they bear no force post optimization. We contrast the total convergence runtime for AD and FD with $h=1\times10^{-9}$ per diagram $\constrainedform$.}
    \label{fig:tensegrity}
\end{figure}

Self-stressed structures are not subjected to external loads and are support-free. 
Here, we model a 2D self-stressed tensegrity wheel such that its perimeter is entirely in tension and the internal spokes in compression (Figure \ref{fig:tensegrity}).
We carry out a parametric study where we progressively increased the number of optimization parameters as we incremented the number of sides on the perimeter of the wheel in steps of size $2^n$, where $n\in\{2,...,8\}$.
Equation \ref{opt_problem} is minimized with the L-BFGS algorithm \cite{nocedal_updatingquasinewton_1980}.
For each configuration, the topological diagram of the wheel comprises $2^{n+1}$ nodes and $2^{n+1} - 0.5(2^n)$ edges, of which $1.5(2^n)$ are deviation edges and the remainder, the edges of the auxiliary trails.
We consider the force in every deviation edge an optimization parameter in $\optimizationvariables$ and test FD with four different step sizes $h$, three orders of magnitude apart, $h\in\{1\times10^{-3},1\times10^{-6},1\times10^{-9},1\times10^{-12}\}$.
All the wheel experiments converge by satisfying the condition $\penaltyoutput\leq\convergencethreshold$ in Equation \ref{eq:optimization_convergence}.

\begin{figure}[!b]
    \captionsetup[subfigure]{justification=centering}
    \centering
    \begin{subfigure}[b]{0.85\columnwidth}
        \centering
        \includegraphics[width=\textwidth]{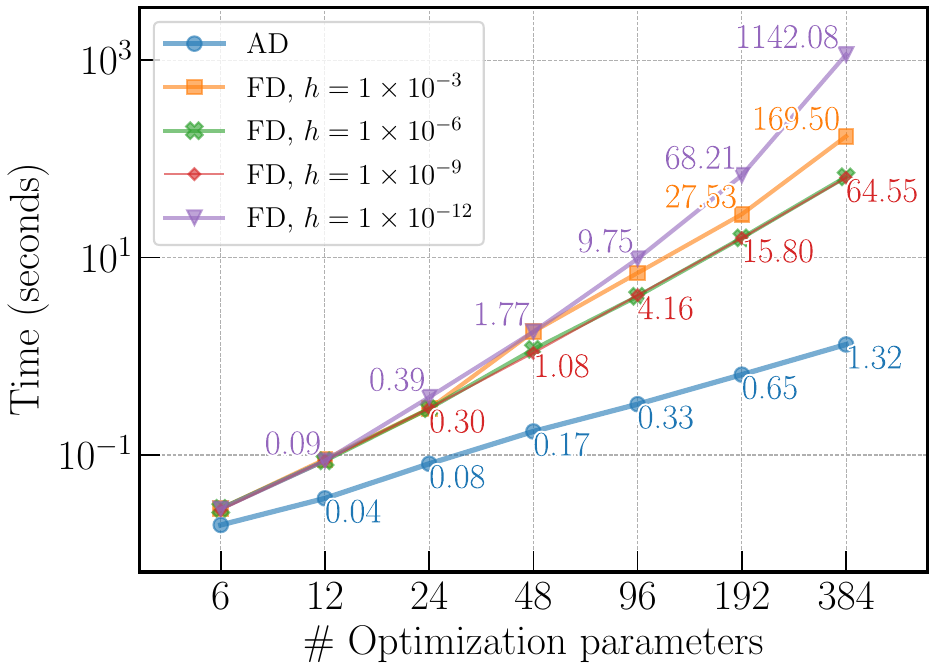}
        \caption{Convergence time}
        \vspace{3mm}
        \label{fig:wheel_time}
    \end{subfigure}
    \begin{subfigure}[b]{0.85\columnwidth}
        \centering
        \includegraphics[width=\textwidth]{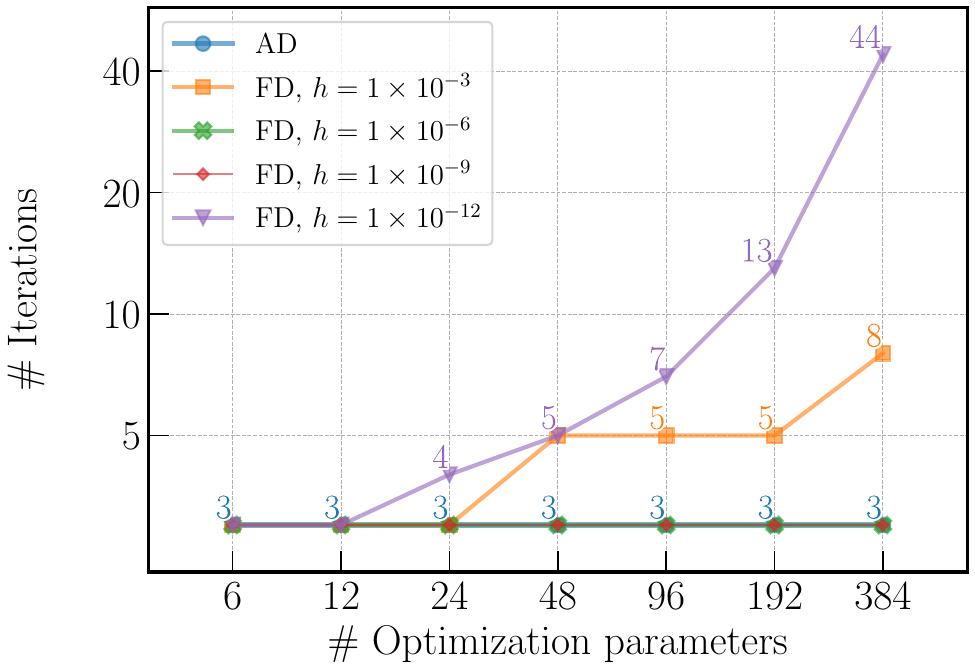}
        \caption{Number of iterations for convergence}
        \label{fig:wheel_iters}
    \end{subfigure}
    \caption{
        Performance comparison between AD and FD to optimize a planar tensegrity wheel. Using FD is computationally more expensive than AD as the number of optimization parameters increases (Figure \ref{fig:wheel_time}).
        Inadequate values of the step size $h$ raise the number of iterations required for convergence and thus extend the convergence runtime (Figure \ref{fig:wheel_iters}).}
    \label{fig:experiments_plots}
\end{figure}

While AD and FD show comparable performance when $n<4$, the total time for convergence with FD surges as the number of optimization parameters increases, irrespective of the step size $h$ (see Figure \ref{fig:wheel_time}).
When the number of parameters is the highest ($n=8$, 384 parameters), form-finding the tensegrity wheel with AD gradients takes only 2.1\% of the time it took to do so with the best FD performance (1.32 vs. 64.55 seconds, when $h=1\times10^{-9}$).
In contrast, the AD convergence time scales linearly with the total number of parameters.

Figures \ref{fig:wheel_time} and \ref{fig:wheel_iters} expose the effect that changing the value of $h$ has on the convergence with FD.
The computation time with FD for a single iteration is equivalent for all the values of $h$ we tested.
However, if $h$ is too large ($h=1\times10^{-3}$) or too small ($h=1\times10^{-12}$), then convergence with FD for this tensegrity structure is slower because the optimizer needs more iterations to reach an optimal solution for $\optimizationvariables$ due to an inaccurate approximation of the gradient $\gradientvalue$.
For example, the optimizer requires 5 and 41 more iterations than AD to solve a tensegrity wheel with $n=8$, when $h=1\times10^{3}$ and $h=1\times10^{-12}$ respectively.

\subsection{Auxiliary trails in 3D}
\label{tree}

\begin{figure*}[t!]
    \centering
    \begin{subfigure}[b]{0.33\textwidth}
        \centering
        \includegraphics[width=\textwidth]{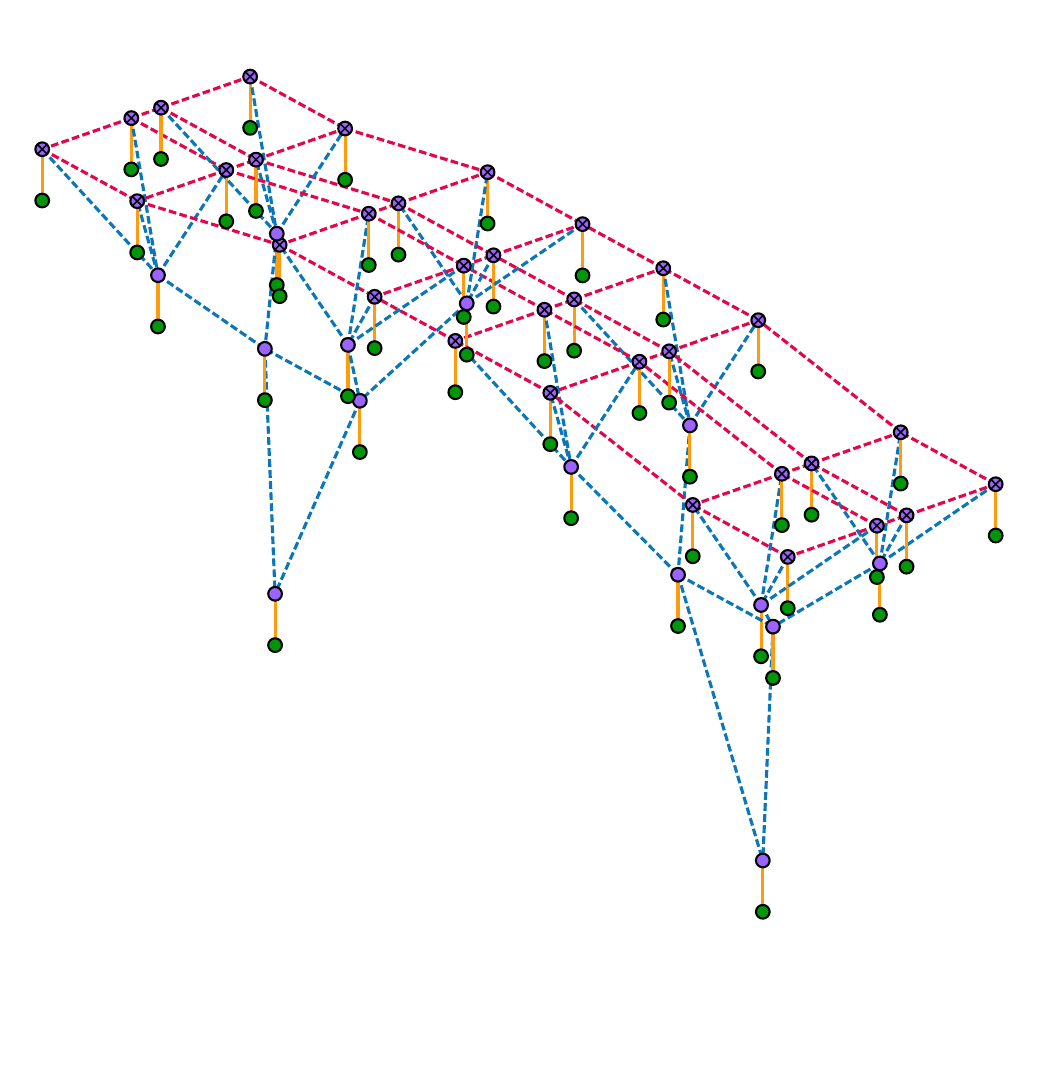}
        \caption{Topology diagram, $\topology$}
    \end{subfigure}
    \begin{subfigure}[b]{0.33\textwidth}
        \centering
        \includegraphics[width=\textwidth]{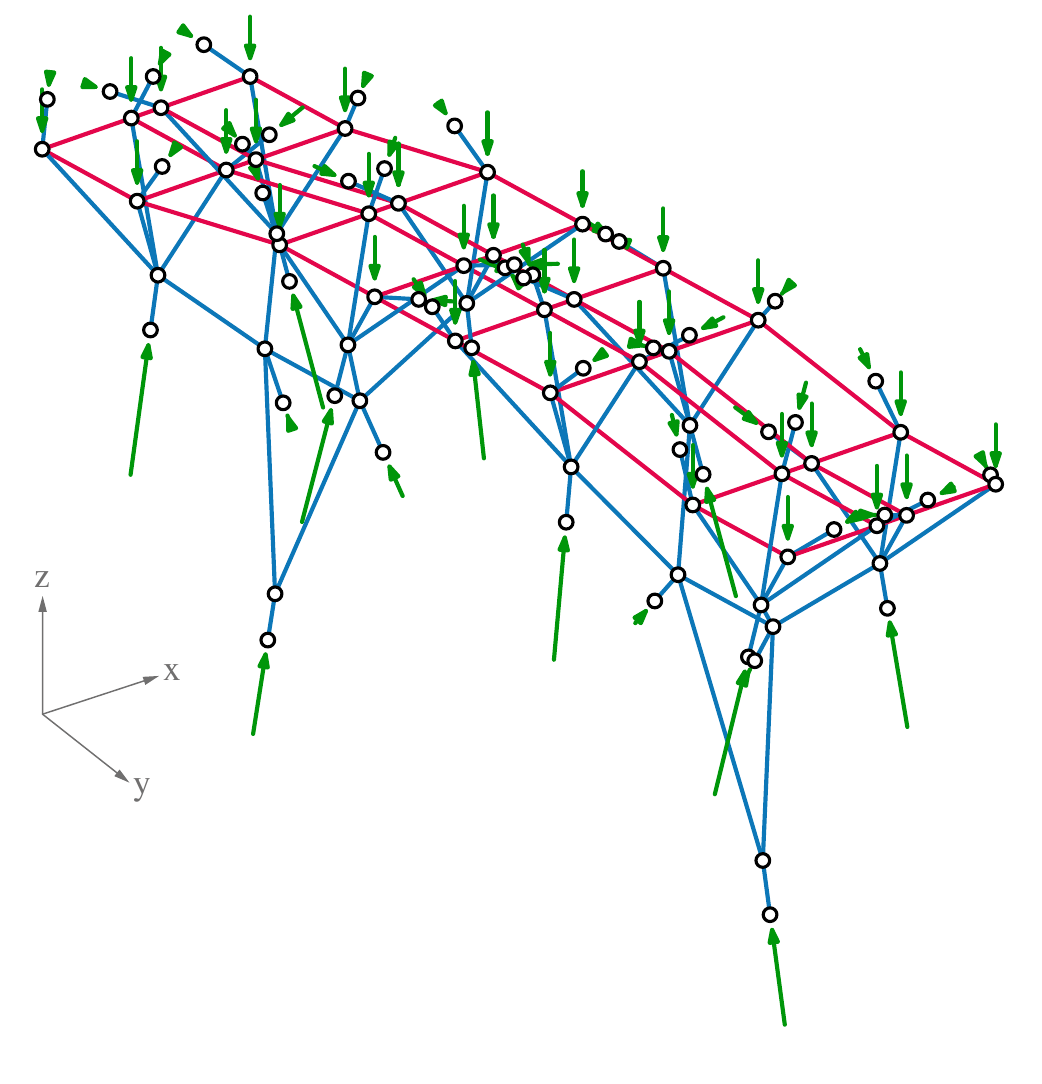}
        \caption{Form diagram, $\form$}
    \end{subfigure}
    \begin{subfigure}[b]{0.33\textwidth}
        \centering
        \includegraphics[width=\textwidth]{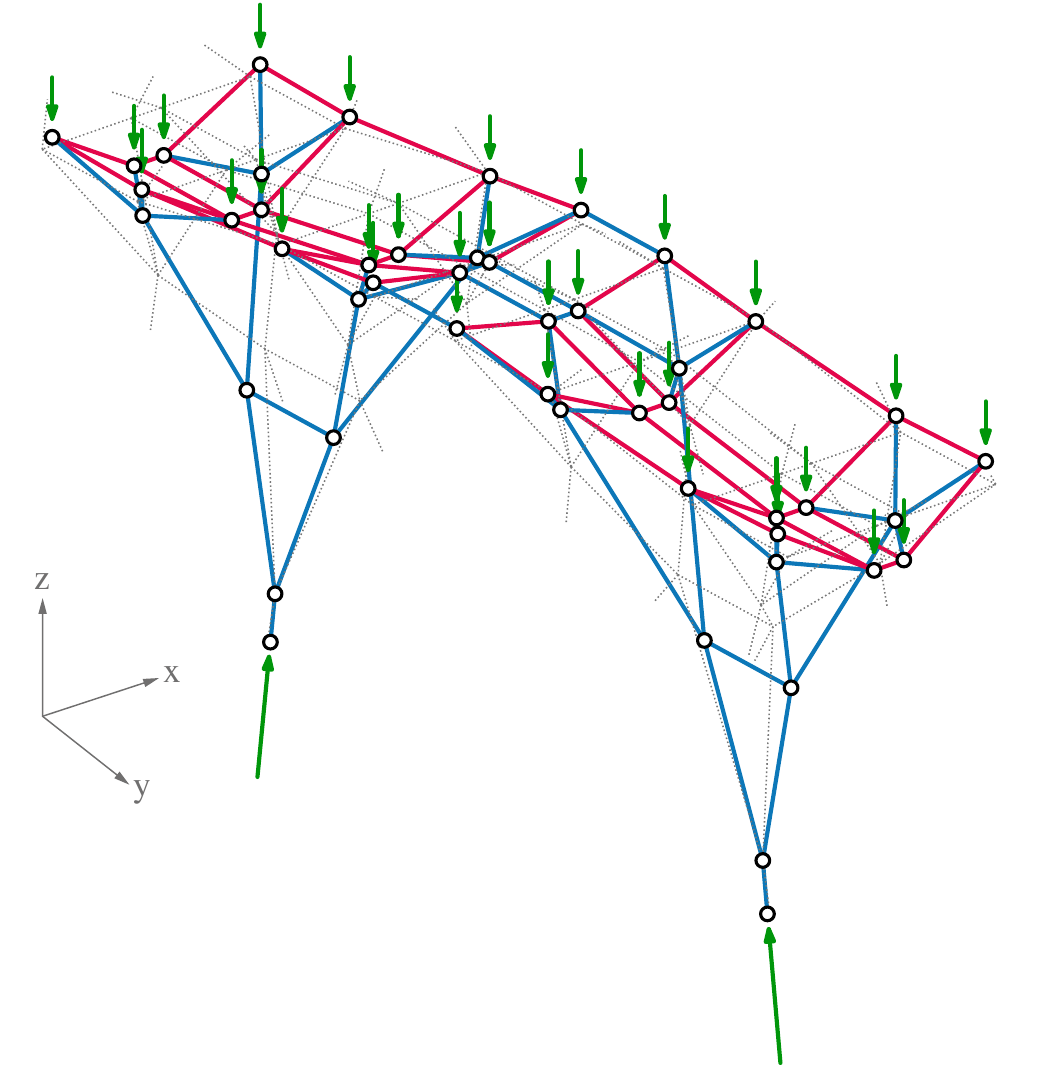}
        \caption{Constrained form diagram, $\constrainedform$}
    \end{subfigure}
    \caption{Tree canopy. The auxiliary trails on the branches take a large portion of the support reaction forces in the form diagram $\form$. After optimizing the position of the nodes, the reaction forces in the constrained form diagram $\constrainedform$ are taken only by the two supports at the base of the structure, while the internal forces in the auxiliary trails vanished.}
    \label{fig:tree_diagrams}
\end{figure*}

We test the addition of auxiliary trails to a three-dimensional tree canopy structure (see Figure \ref{fig:tree_diagrams}). 
We observe how the performance of AD and FD differs as the choice of optimization algorithm changed. 
We model the initial tree structure with 46 nodes and 72 deviation edges.
The total number of nodes and edges in the topological diagram doubles after inserting the auxiliary trails.
We apply a point load $\nodeload=[0, 0, -0.5]$ to the nodes at the top layer of the structure.

To minimize the forces in the auxiliary trails, we define 186 optimization parameters. 
These parameters consist of the positions of the origin nodes, which are allowed to translate only in the Y and Z Cartesian directions, and the force magnitude in every deviation edge. 
We test three different gradient-based optimization algorithms: L-BFGS \cite{nocedal_updatingquasinewton_1980}, SLSQP \cite{kraft_algorithm733_1994}, and AUGLAG \cite{conn_globallyconvergent_1991} and run them for a maximum of $u^{\text{max}}=200$ optimization iterations.
The FD step sizes we discuss for this structure are $h\in\{1\times10^{-6},1\times10^{-9}\}$.

\begin{figure}[!b]
    \centering
    \includegraphics[width=0.85\columnwidth]{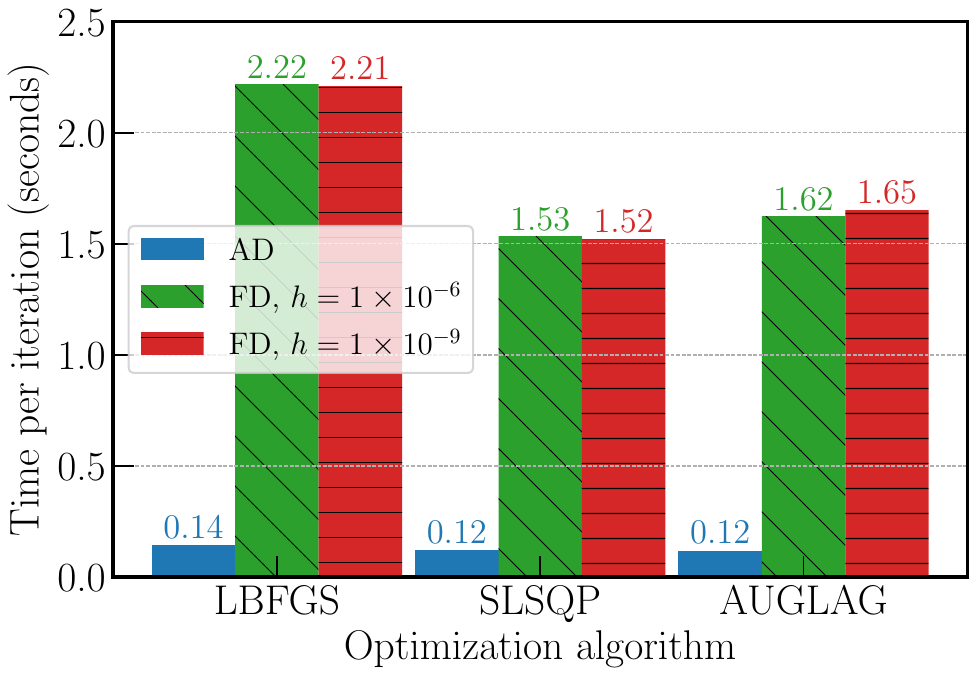}
    \caption{Tree canopy. Convergence with AD is at least 10 times faster per iteration than FD. This disparity in computational performance is consistent for this constrained form-finding problem regardless of the optimization algorithm.}
    \label{fig:tree_solvers}
\end{figure}

In Figure \ref{fig:tree_solvers}, we report the elapsed time per iteration that the optimization algorithms took to converge by meeting the condition $\penaltyoutput\leq\convergencethreshold$ in Equation \ref{eq:optimization_convergence} with $\convergencethreshold=1\times10^{-6}$.
We calculate the time per iteration by dividing the convergence runtime over the total number of iterations incurred.
This ratio is nearly equal for both FD step sizes across the three optimization algorithms we assessed.
Minimizing the forces in the auxiliary trails of the tree structure is at least 10 times faster per iteration using AD than with FD.
The optimization time with AD is at most 0.14 seconds per iteration with L-BFGS, whereas this value rises to 1.52 seconds with FD and SLSQP.

Among the three optimization algorithms we test, L-BFGS takes the least number of iterations to solve this constrained form-finding problem for both differentiation schemes, requiring 78 and 120 iterations to converge for AD and FD, respectively.
Moreover, the minimization of Equation \ref{opt_problem} with AD is at least one order of magnitude faster for the 2D tensegrity structure described in Section \ref{wheel} than it is with the tree structure presented here, despite the size of the two constrained form-finding problems is similar and the optimization algorithm is the same: 192 parameters and a convergence runtime of 0.65 seconds for the tensegrity versus 182 parameters and 11.15 seconds for the tree.

A plausible reason for this discrepancy is that, for the tree structure, both the node positions and the deviation edge forces are set as optimization parameters, whereas for the spoke wheel the optimization parameters only contemplate the forces in the deviation edges.
The minimization of the auxiliary trail forces utilizing both the node positions and the internal forces of the structure is a non-linear problem that can be computationally more expensive to solve.

\begin{figure}[t!]  
    \centering
    \includegraphics[width=\columnwidth]{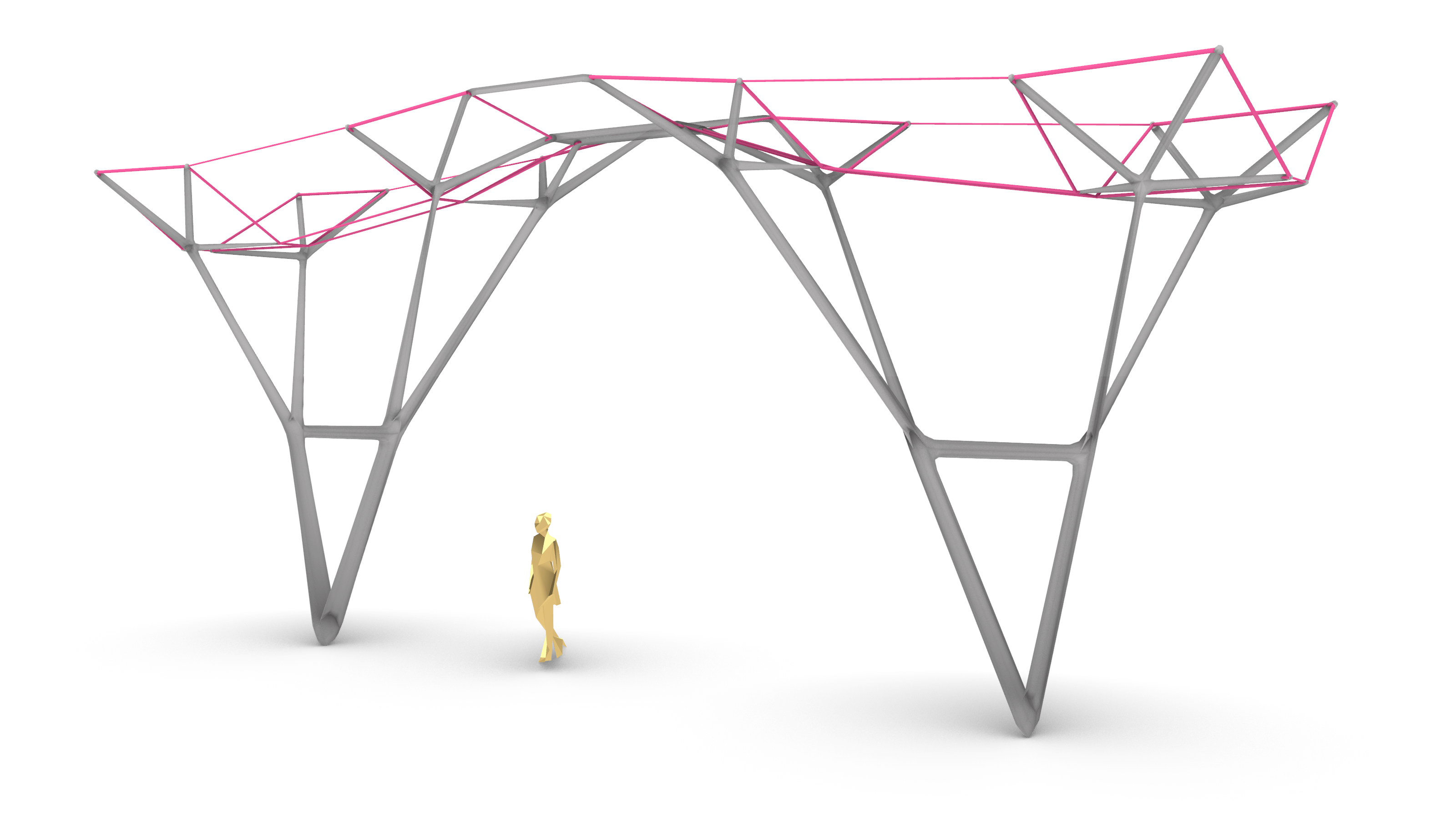}
    \includegraphics[width=\columnwidth]{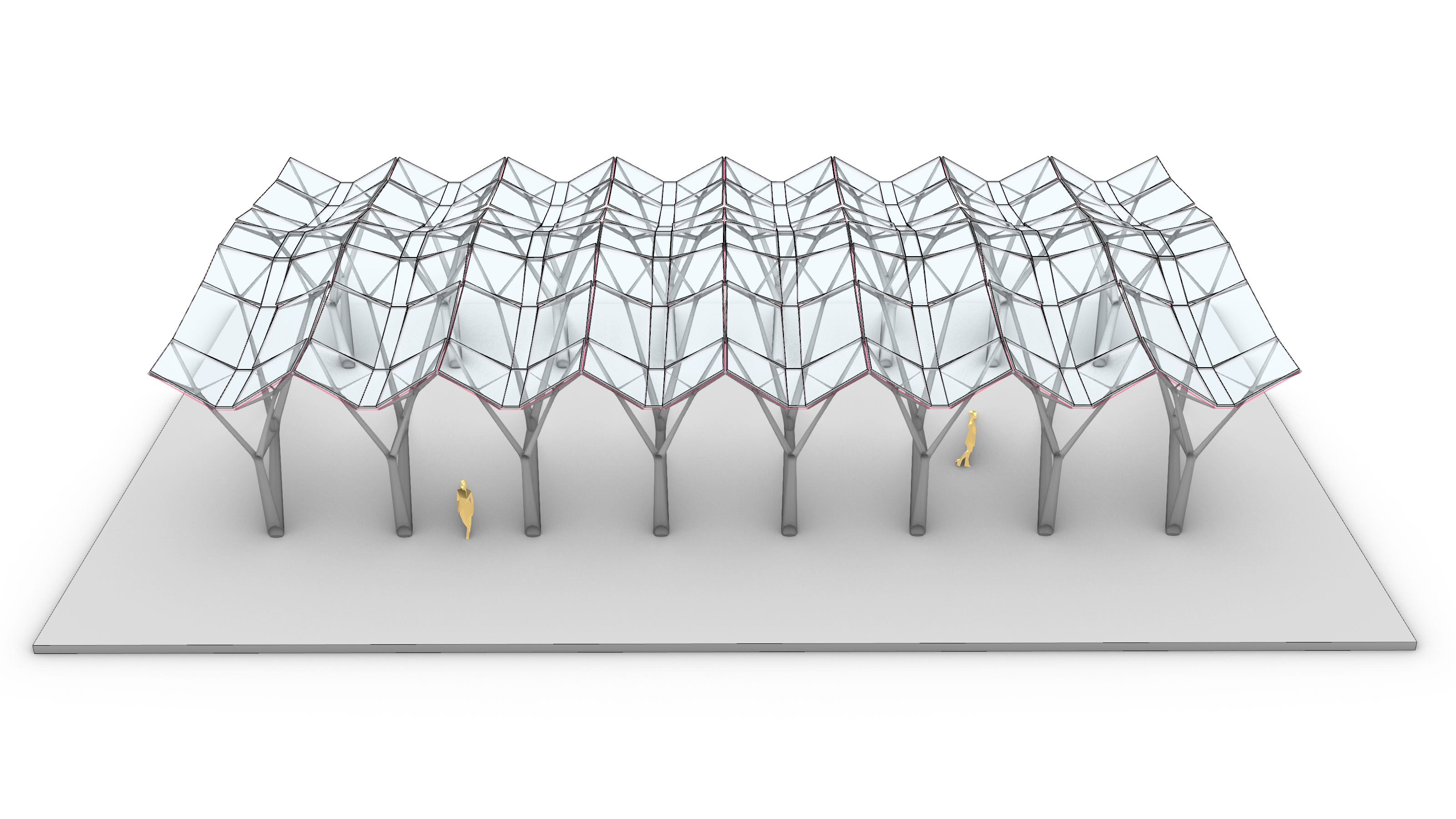}\vspace{-1cm}
    \caption{Architectural vision for the constrained form diagram $\constrainedform$ shown in Figure \ref{fig:tree_diagrams}. The form-found tree canopy is repeated sequentially to create a colonnade of load-bearing trees.}
    \label{fig:tree_colonnade}
\end{figure}

\subsection{Auxiliary trails in 3D and additional constraints}
\label{bridge}

\begin{figure}[t!]
    \captionsetup[subfigure]{justification=centering}
    \centering
    \begin{subfigure}[b]{\columnwidth}
        \centering
        \includegraphics[trim={0 0 0 20mm},clip,width=\textwidth]{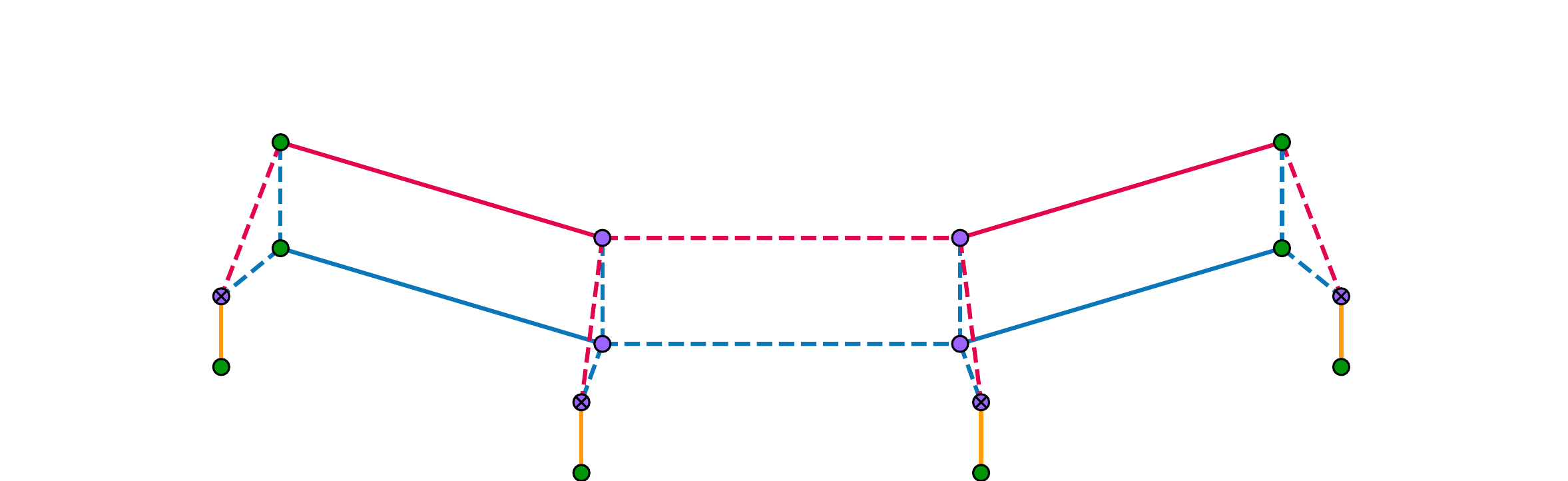}
        \caption{Topology diagram $\topology$ (4 hangers)}
    \end{subfigure}
    \begin{subfigure}[b]{\columnwidth}
        \centering
        \begin{subfigure}[b]{\columnwidth}
            \centering
            \vspace{-3mm}
            \includegraphics[width=\textwidth]{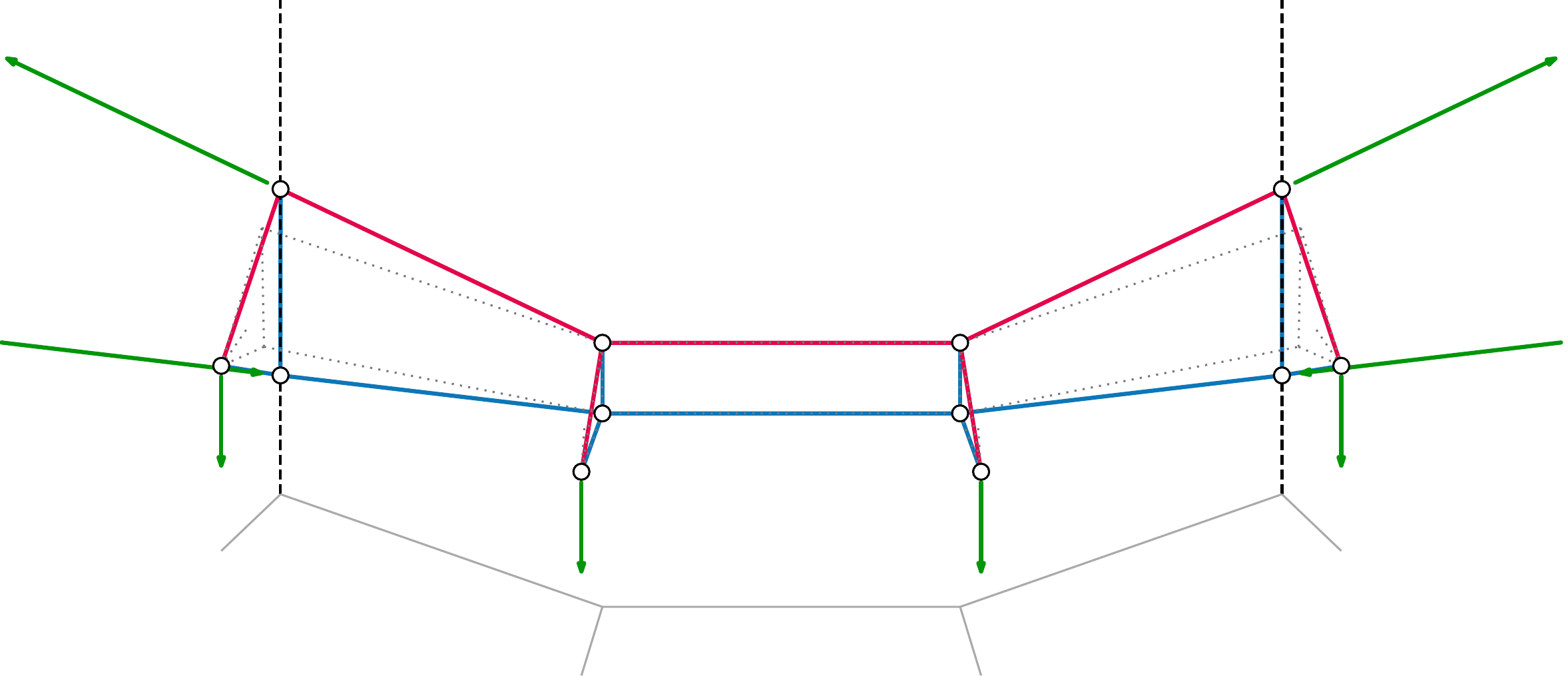}
            \caption*{18 parameters (4 hangers)\\AD: 0.5 sec / FD: 1.2 sec}
        \end{subfigure}
        \begin{subfigure}[b]{\columnwidth}
            \centering
            \vspace{-6mm}
            \includegraphics[width=\textwidth]{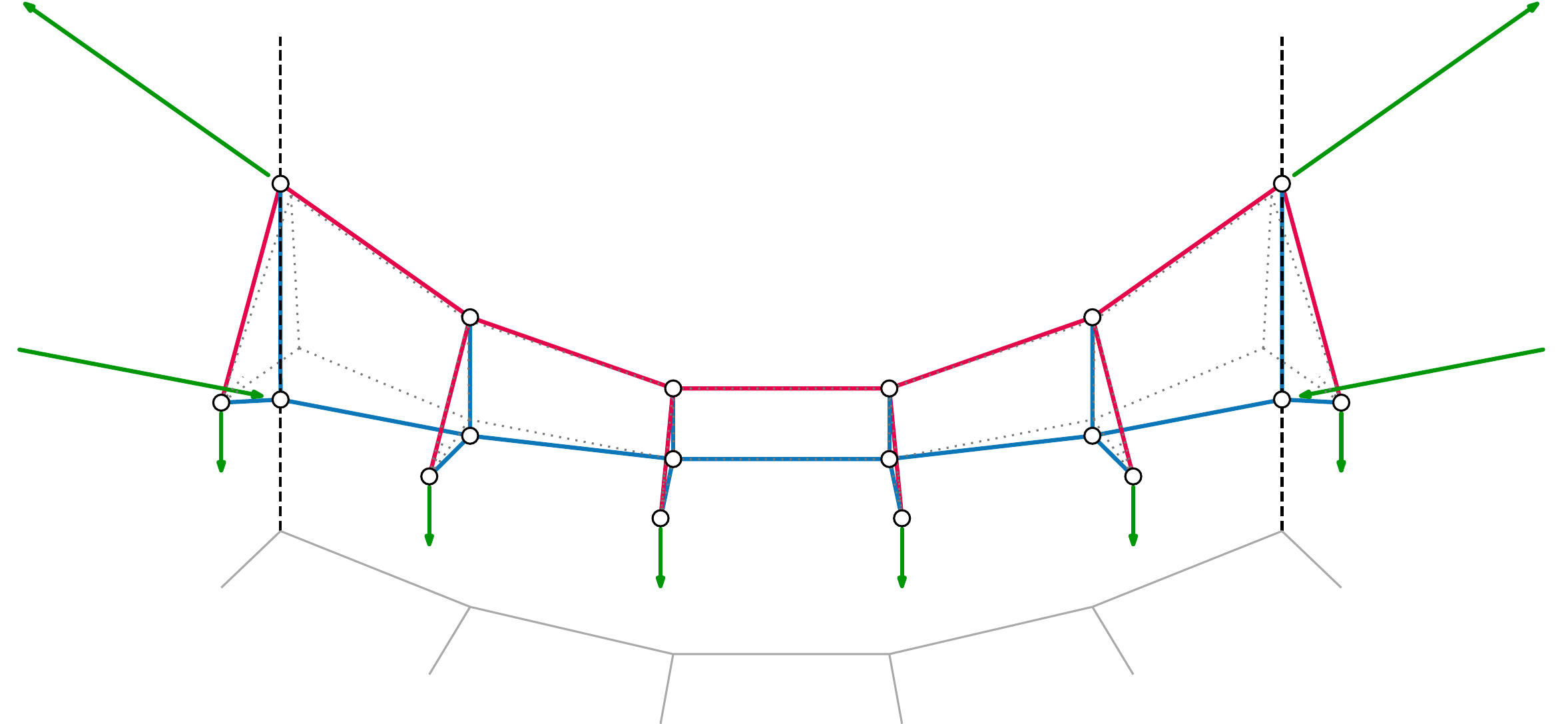}
            \caption*{24 parameters (6 hangers)\\AD 1.5 sec / FD: 3.3 sec}
        \end{subfigure}
        \begin{subfigure}[b]{\columnwidth}
            \centering
            \vspace{-6mm}
            \includegraphics[width=\textwidth]{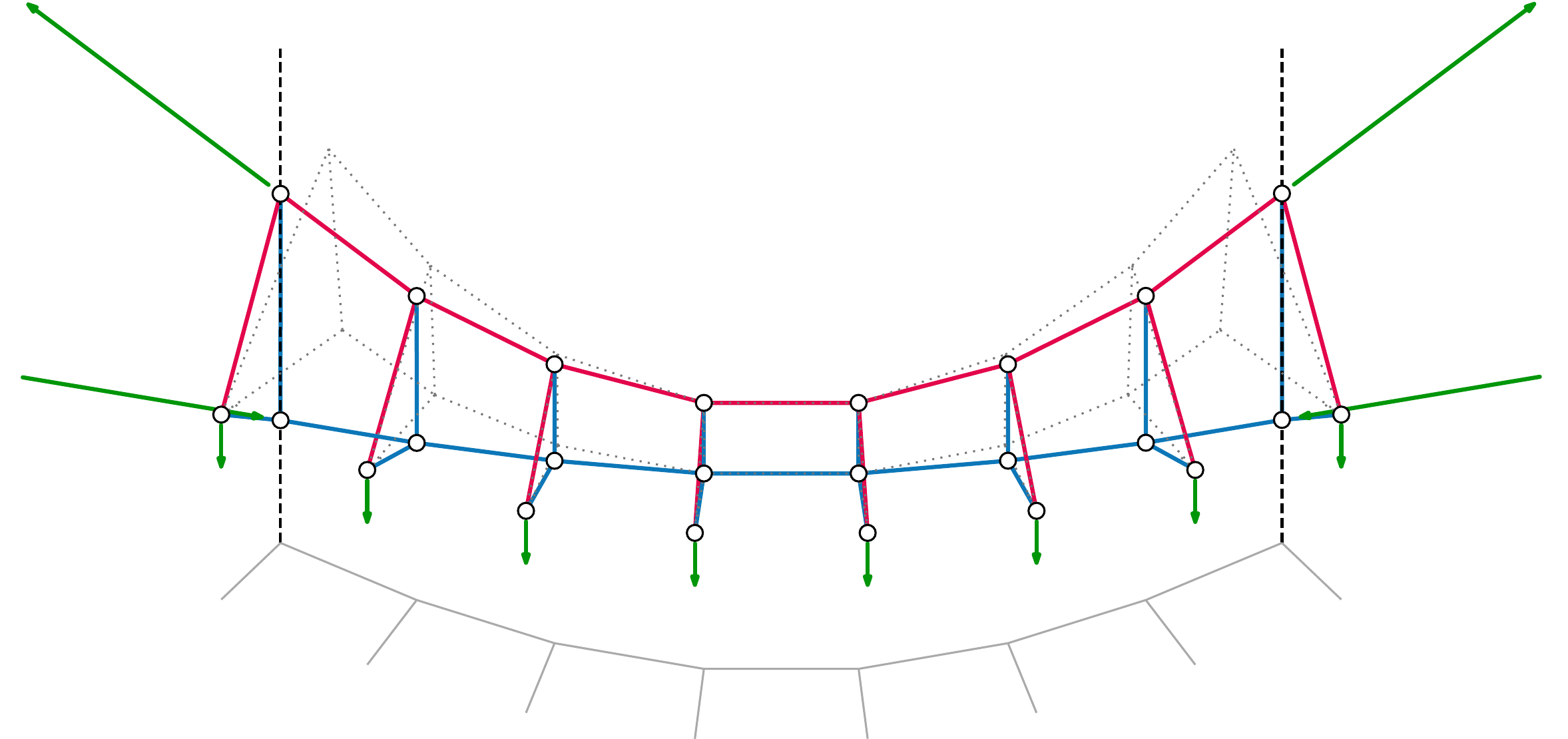}
            \caption*{30 parameters (8 hangers)\\AD: 3.1 sec / FD: 7.7 sec}
        \end{subfigure}
        \begin{subfigure}[b]{\columnwidth}
            \centering
            \vspace{-4mm}
            \includegraphics[width=\textwidth]{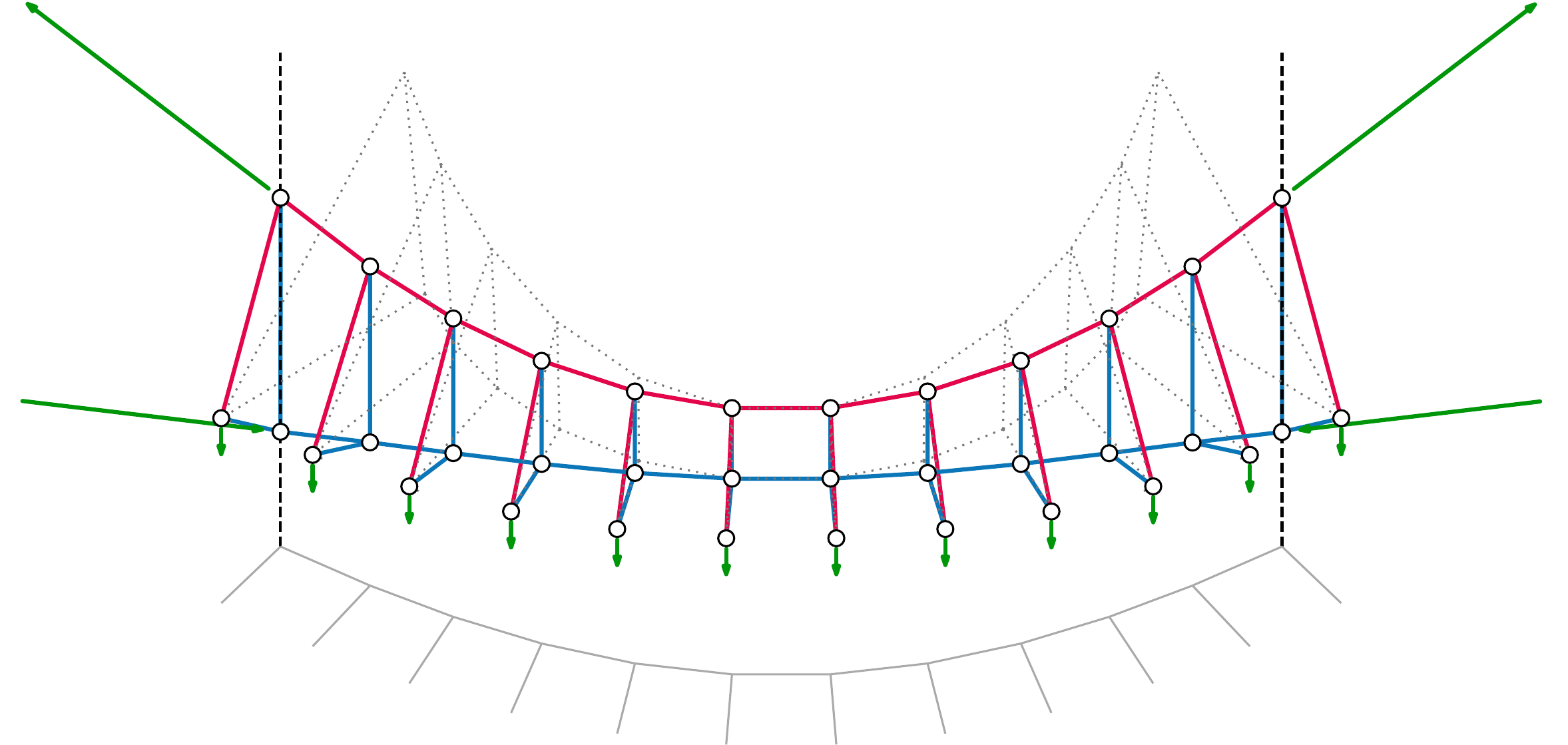}
            \caption*{42 parameters (12 hangers)\\AD: 7.3 sec / FD: 23.0 sec}
        \end{subfigure}
        \caption{Constrained form diagrams $\constrainedform$}
    \end{subfigure}
    \caption{Sensitivity analysis of a 3D bridge curved on plan.
    The form diagrams $\form$ obtained after the first run of the CEM form-finding algorithm are drawn in the background with dotted lines.
    The forces in the auxiliary trails at the tip of the hangers have been minimized and the bridge endpoints pulled to the dashed vertical lines in black.
    The solution to this constrained form-finding problem is consistently faster with AD.}
    \label{fig:bridge_sensitivity}
\end{figure}

We combine force and geometric constraints to steer the form-finding of a bridge with no intermediary supports (see Figure \ref{fig:bridge_sensitivity}). 
The bridge is curved on plan in the longitudinal direction and a series of triangular hangers in its transversal direction takes a uniformly distributed line load produced by an eccentric runway.
We convert this torsional into point loads via tributary lengths and apply them to the tip of the hangers.
Unlike the spoke wheel and the tree structure discussed in Sections \ref{wheel} and \ref{tree}, we model the topological diagram $\topology$ of the bridge using a hybrid strategy: we append auxiliary trails only to the nodes to the tip of the hangers because we assign all the other nodes in $\topology$ to a standard trail on the longitudinal chords of the bridge.

The geometric constraint for this structure is to pull the support nodes towards two predetermined vertical line rays located at either extreme of the bridge, in addition to vanishing the forces in the auxiliary trails.
The rationale behind this constraint is to arrive at a form in static equilibrium subject to a limited range of locations to anchor the bridge abutments.
The forces in all the deviation edges in $\topology$ are considered optimization parameters, in addition to the length of the four trail edges connected to a support node.

\begin{figure}[!t]
    \captionsetup[subfigure]{justification=centering}
    \centering
    \begin{subfigure}[b]{0.85\columnwidth}
        \centering
        \includegraphics[width=\textwidth]{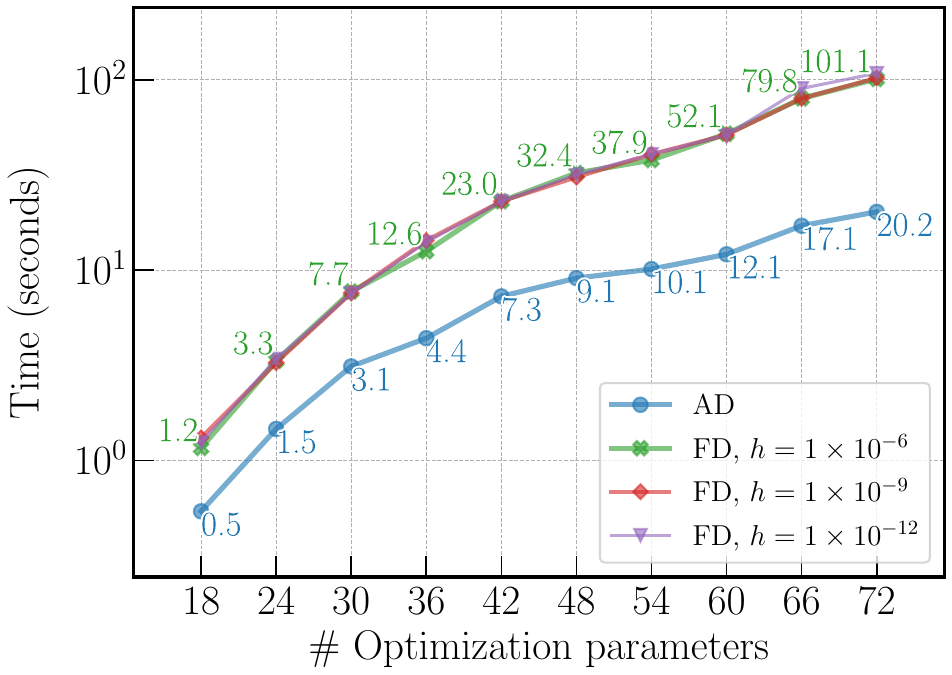}
        \caption{Convergence time}
        \vspace{3mm}
        \label{fig:bridge_time}
    \end{subfigure}
    \begin{subfigure}[b]{0.85\columnwidth}
        \centering
        \includegraphics[width=\textwidth]{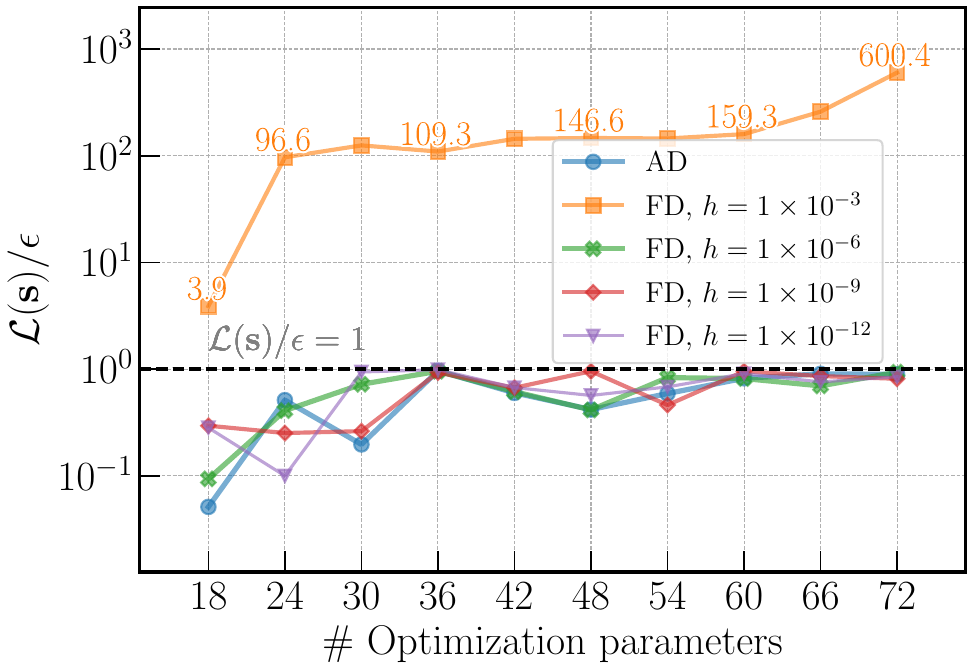}
        \caption{Convergence rate}
        \label{fig:bridge_convergence}
    \end{subfigure}
    \caption{Curved bridge. Optimizing the bridge with FD is consistently more expensive than with AD regardless of the step size $h$. (Figure \ref{fig:bridge_time}). The optimizer misses the optimization convergence threshold $\convergencethreshold=1\times 10^{-6}$ by at least two orders of magnitude with FD and $h=1\times 10^{-3}$ (Figure \ref{fig:bridge_convergence}).}
    \label{fig:bridge_experiments}
\end{figure}

We do a sensitivity analysis and compare AD and FD by monotonically increasing the number of hangers, from 4 to 22 in even steps, and then solving the resulting bridge constrained form-finding problem with SLSQP \cite{kraft_algorithm733_1994}.
The number of optimization parameters ranges from 18, when the number of triangular hangers is the smallest, to 72 when it is the greatest.
Our goal of is to estimate the time this constrained form-finding problem would take to converge to $\penaltyoutput\leq\convergencethreshold$, where $\convergencethreshold=1\times10^{-6}$, when we add auxiliary trails only to a portion of the nodes in $\topology$ of the bridge.
We set the step size for FD to $h\in\{1\times10^{-3}, 1\times10^{-6},1\times10^{-9},1\times10^{-12}\}$ and we restrict the number of optimization iterations to $\upsilon^{\text{max}}=100$ for both differentiation methods, AD and FD.

Figure \ref{fig:bridge_time} shows that the time for convergence with FD is nearly the same for three different step sizes $h$ across all experiments.
The optimizer did not converge with $h=1\times10^{-3}$ for this structure.
This is different from the observations we make after studying a planar tensegrity in Section \ref{wheel}, where FD with a step size of $h=1\times10^{-3}$ converges and $h=1\times10^{-12}$ extends the convergence runtime of the optimizer with FD.
This finding illustrates that the impact of $h$ on the quality of the gradient approximation is problem dependent.

Figure \ref{fig:bridge_convergence} provides insight into the inadequacy of $h=1\times10^{-3}$.
The final value of the objective function $\penaltyoutput$ is, on average, two orders of magnitude higher than the desired optimization convergence threshold $\convergencethreshold$.
In contrast, AD and FD with the three other step sizes $h$ meet the target value of $\convergencethreshold$ within the iteration budget $\upsilon^{\text{max}}=100$ since they reach $\penaltyoutput/\convergencethreshold\leq1$.
Nevertheless, optimizing the bridge is consistently more expedite with AD: convergence with a AD is 2.4 and 5.0 times faster when the number of parameters is the smallest (22) and the largest (72) respectively.

\begin{figure*}[b!]
    \centering
    \begin{subfigure}[b]{\textwidth}
        \centering
        \includegraphics[width=0.95\textwidth]{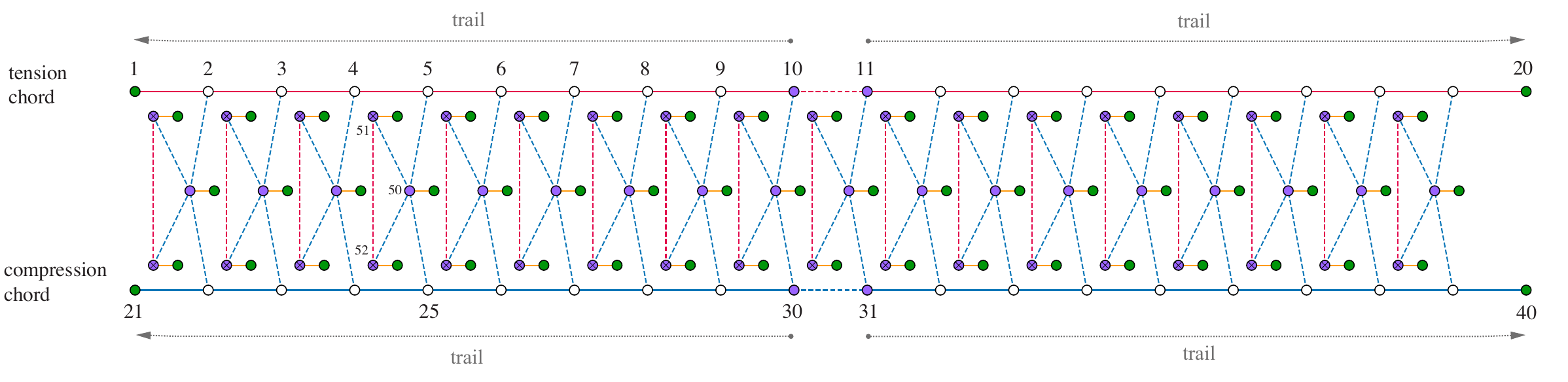}
        \caption{
        Topology diagram with auxiliary trails, $\topology$.
        The tension and compression chords of the staircase are represented by two trails and one deviation edge each.
        Edges $\deviationedge_{5, 50}, \deviationedge_{25, 50}, \deviationedge_{52, 50}, \deviationedge_{52, 50}$ model one of the 18 compression ribs and represents $\deviationedge_{51, 52}$ its matching tension tie.
        The design load was of 1 kN per step was applied as $\nodeload=[0,0,-0.5]\text{ kN}$ to the end-nodes on the tension ties (e.g. to nodes $\nodeorigin_{51}, \nodeorigin_{52}$).}
        \label{fig:staircase_topology}
    \end{subfigure}
    \bigskip
    \begin{subfigure}[b]{0.45\textwidth}
        \centering
        \includegraphics[width=\columnwidth, trim={1cm 0 1cm 0}]{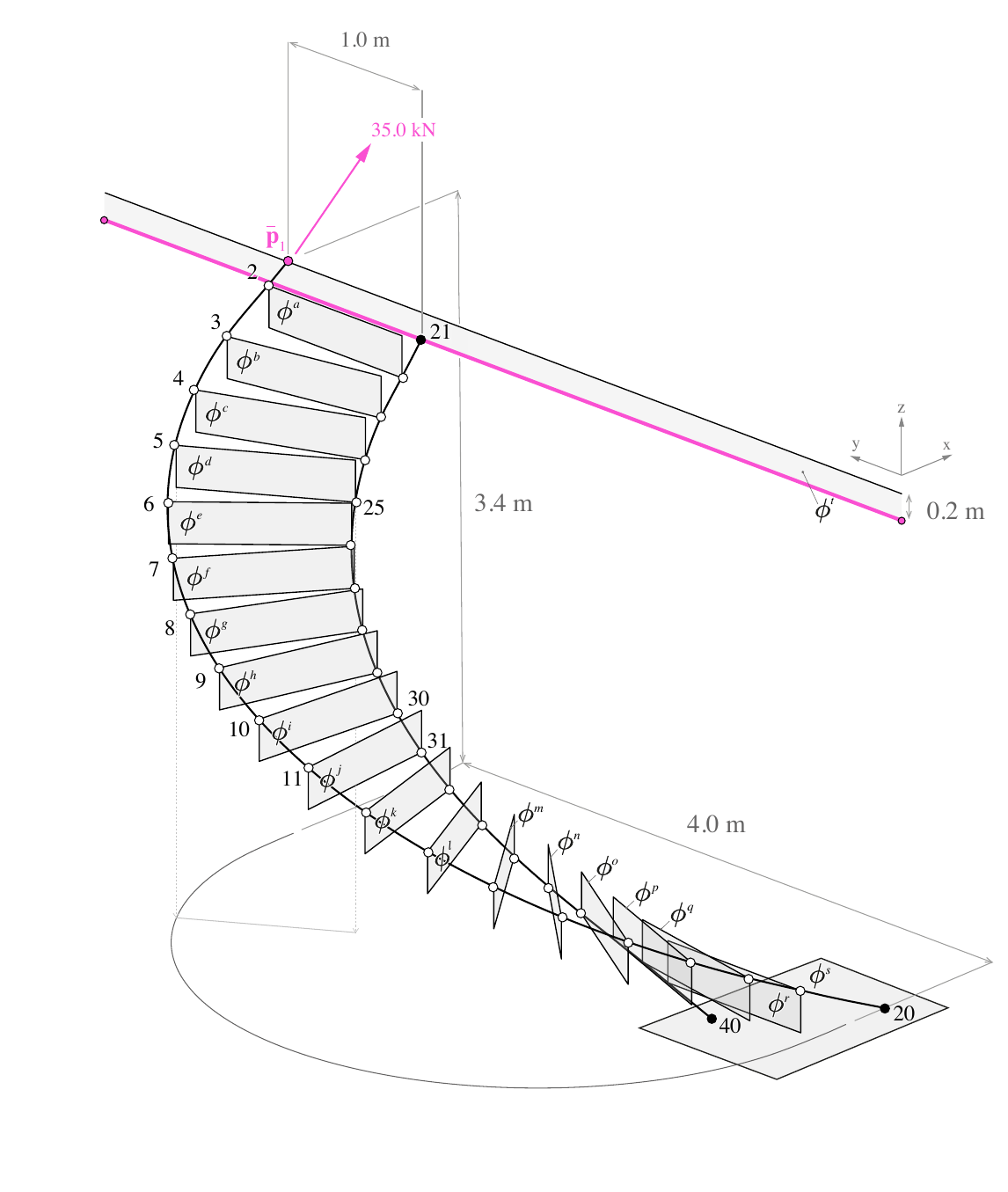}
        \caption{Design constraints}
        \label{fig:staircase_constraints}
    \end{subfigure}
    \begin{subfigure}[b]{0.45\textwidth}
        \centering
        \includegraphics[width=\columnwidth, trim={1cm 0 1cm 0}]{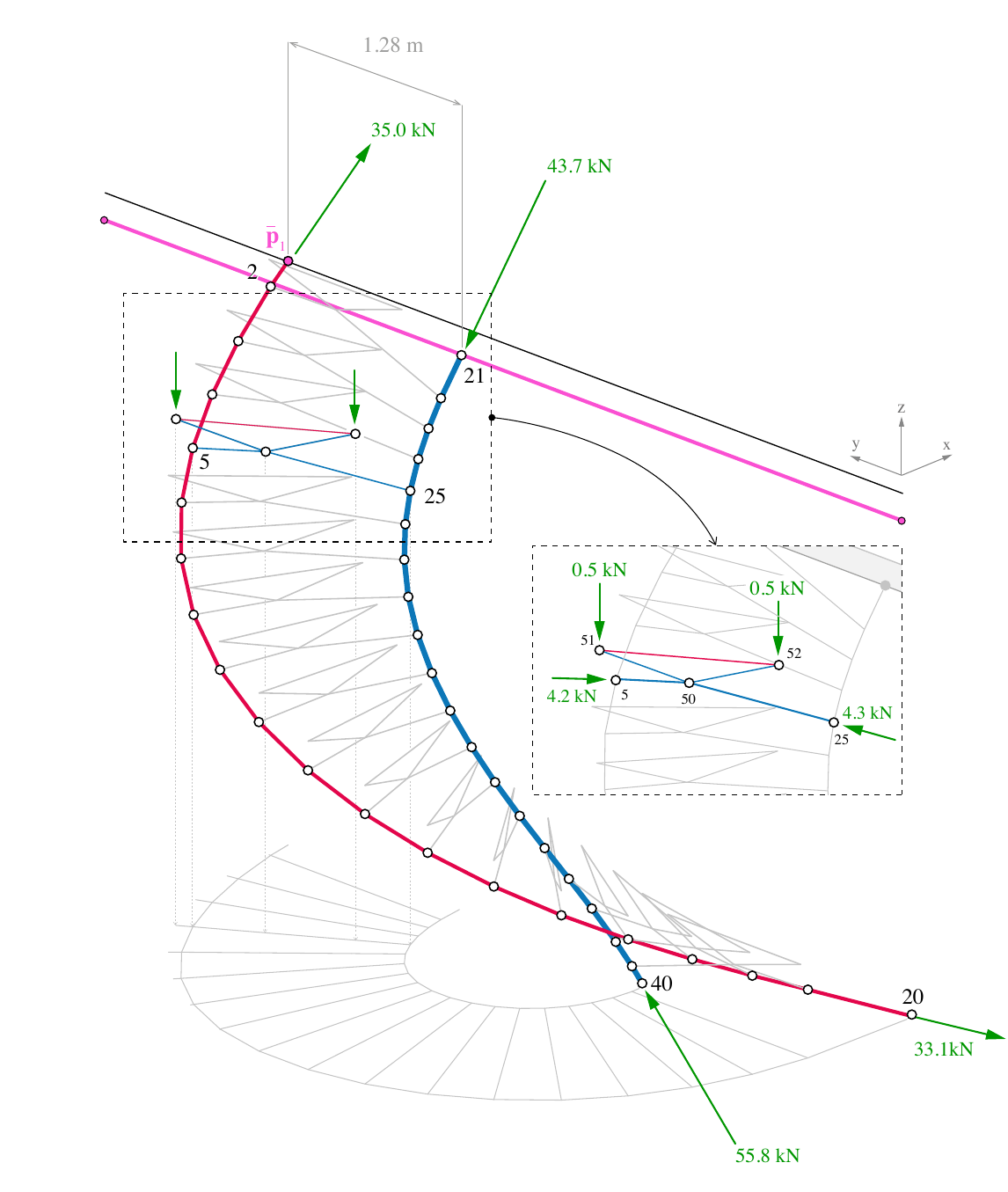}
        \caption{Constrained form diagram, $\constrainedform$}
        \label{fig:staircase_form}
    \end{subfigure}
    \vspace{-4mm}
    \caption{Application of the extended CEM framework to design the load-bearing structure of a spiral staircase.
    After constructing a topology diagram $\topology$ (Figure \ref{fig:staircase_topology}), a constrained form diagram $\constrainedform$ (Figure \ref{fig:staircase_form}) that complies with a priori structural and fabrication constraints is computed using AD.
    The design constraints input to Equation \ref{opt_problem} in addition to the minimization of the forces in the auxiliary trails are shown in pink in Figure \ref{fig:staircase_constraints} and comprise: i) restraining the position of node $\nodesupport_1$ to target position $\bar{\nodeposition}_1=[0,4,3.4]$, ii) fixing the magnitude of support force $\supportforce_1$ to 35 kN; and iii) constraining $\nodeposition_{21}$ to lie on a continuous horizontal line 0.2 meters below $\bar{\nodeposition}_1$.
    We also restrict via the CEM form-finding algorithm the position of every pair of unsupported nodes on the chords to the plane defined by the nodes of the funicular rib they connect to (e.g. $\node_{5}$ and $\node_{25}$ to $\trailedgeplane^d$), the position that of support nodes $\nodesupport_{20},\nodesupport_{40}$ to the ground floor plane $\trailedgeplane^s$ and that of nodes $\nodesupport_{1},\nodesupport_{21}$ to the slab plane $\trailedgeplane^t$}
\end{figure*}

\section{Case study}
\label{casestudy}

We illustrate the potential of the extended CEM framework to support designers in practical structural design problems, especially during the conceptual design stage.

\subsection{Design task}

We design the load-bearing structure of a spiral staircase subjected to the design constraints listed in Section \ref{design_constraints}.
The external perimeter of the staircase follows a semicircle on-plan with a diameter of 4 meters (Figure \ref{fig:staircase_constraints}).
The staircase is planned to connect the ground floor to the mezzanine slab with a single run of 18 equidistant steps.
Every step is 1 meter wide and perpendicular to the semicircle.
The mezzanine slab is 3.4 meters above the ground floor and the design load is of 1 kN per step.

Inspired by the Fourth Bridge over the Grand Canal in Venice \cite{zordan_fourthbridge_2010}, we use the extended CEM framework to form-find a spiraling truss-like structure for the staircase.
The structure is imagined to carry the applied loads via cross-shaped ribs suspended on two curving chords.
The chords are initially proposed to be 1 meter apart from each other, running parallel to the run of stairs and to be anchored at their intersections with the ground floor and the shallow side of the mezzanine slab.

The static equilibrium concept is to first use the tread in every step of the staircase as a tension element that tied the upper half of the compression rib underneath (see callout rectangle in Figure \ref{fig:staircase_form}).
Next, the goal is to resist the torsional effects produced by the forces coming from the ribs by coupling the two chords as a pair of tension-compression rings.

\subsection{Topology diagram}
Figure \ref{fig:staircase_topology} displays the topology diagram $\topology$ we build to form-find the staircase.
We represent each of the chords with two trails connected by a single deviation edge at the middle since the chords are the two main paths for load transfer towards the supports.
As secondary load-transfer elements, we model the 18 ribs and ties with deviation edges and auxiliary trails.

\subsection{Design constraints}
\label{design_constraints}

The form-found shape of the staircase structure has to conform to a number of design requirements in addition to the minimization of the forces in the 54 auxiliary trails in $\topology$.
We show these graphically in Figure \ref{fig:staircase_constraints}.

The side of the mezzanine slab where the chords have to be anchored has a maximum pull-out force capacity of 35 kN at position $\bar{\nodeposition}_1=[0,4, 3.4]$.
As a result, the position of support node $\nodesupport_1$ in the tension chord must coincide with $\bar{\nodeposition}_1$, and the target absolute force magnitude passing through edge $\edge_{1,2}$ has to be constrained to $\bar{\edgeforce}_{1,2}=35\,\text{kN}$.
The position of the support node on the compression chord $\nodesupport_{21}$ is restricted to slide on a horizontal line 0.20 meters below $\bar{\nodeposition}_1$, parallel to the bottom edge of the soffit of the mezzanine slab.
We set these design requirements as optimization constraints.

Additionally, we define a sequence of planes to constrain, with the CEM form-finding algorithm, the position of the nodes on the chords to the plane formed by the upper portion of the rib they connect to (see Section \ref{sequential}).
For example, nodes $\node_{5}$ and $\node_{25}$ in $\topology$ should lie on the plane formed by $\nodeorigin_{50}, \nodeorigin_{51}, \nodeorigin_{52}$, which is plane $\trailedgeplane^{d}$ in Figure \ref{fig:staircase_constraints}.
Numerically, plane $\trailedgeplane^{d}$ corresponds to the intersection planes $\trailedgeplane_{6,5}$ and $\trailedgeplane_{26,25}$ of trail edges $\trailedge_{6,5}$ and $\trailedge_{26,25}$ in $\topology$, respectively.
The reasoning behind this planarity constraint is to enable the fabrication of the ribs from flat sheets of material.
Similarly, we pull the positions of the bottom support node per chord to the ground floor plane $\trailedgeplane^{s}$ and that of the support nodes at the top to $\trailedgeplane^{t}$ to explicitly restrict the feasible range of positions of these nodes can take during the optimization process.
Plane $\trailedgeplane^{s}$ is described by base point $\mathbf{p}^{\phi^{s}}=[0,0,0]$ and normal $\mathbf{n}^{\phi^{s}}=[0,0,1]$, whereas plane $\trailedgeplane^{t}$ is defined by $\mathbf{p}^{\phi^{t}}=\bar{\nodeposition}_1$ and $\mathbf{n}^{\phi^{t}}=[1,0,0]$. 

\subsection{Constrained form diagram}

We parametrize this constrained form-finding problem by setting the absolute force magnitude in all the deviation edges $\deviationedgeforce$ as entries in the vector of optimization parameters $\optimizationvariables$.
We also allow the position of the origin nodes on the chords $\nodeorigin_{10}, \nodeorigin_{11}, \nodeorigin_{30}, \nodeorigin_{31}$ to translate vertically.
The resulting optimization problem is minimized with L-BFGS \cite{nocedal_updatingquasinewton_1980}.

We show the resulting constrained form diagram $\constrainedform$ in Figure \ref{fig:staircase_form}.
The output values of the gradient $\gradientvalue$ and of the objective function $\penaltyoutput$ reaches the optimization convergence threshold $\convergencethreshold=1\times10^{-6}$, implying that the generated $\constrainedform$ for the staircase is satisfactory, meeting all the imposed design constraints.
We can observe that some of the trade-offs made to solve this constrained form-finding task are that the initial distance between the chord supports on the mezzanine slab increased from 1 meter to 1.28 meters and that the absolute magnitude of the reaction forces at the compression chord supports are about 25\% and 60\% higher than that at node $\nodesupport_1$ post-optimization.

\begin{figure}[!t]
    \centering
    \includegraphics[width=\columnwidth]{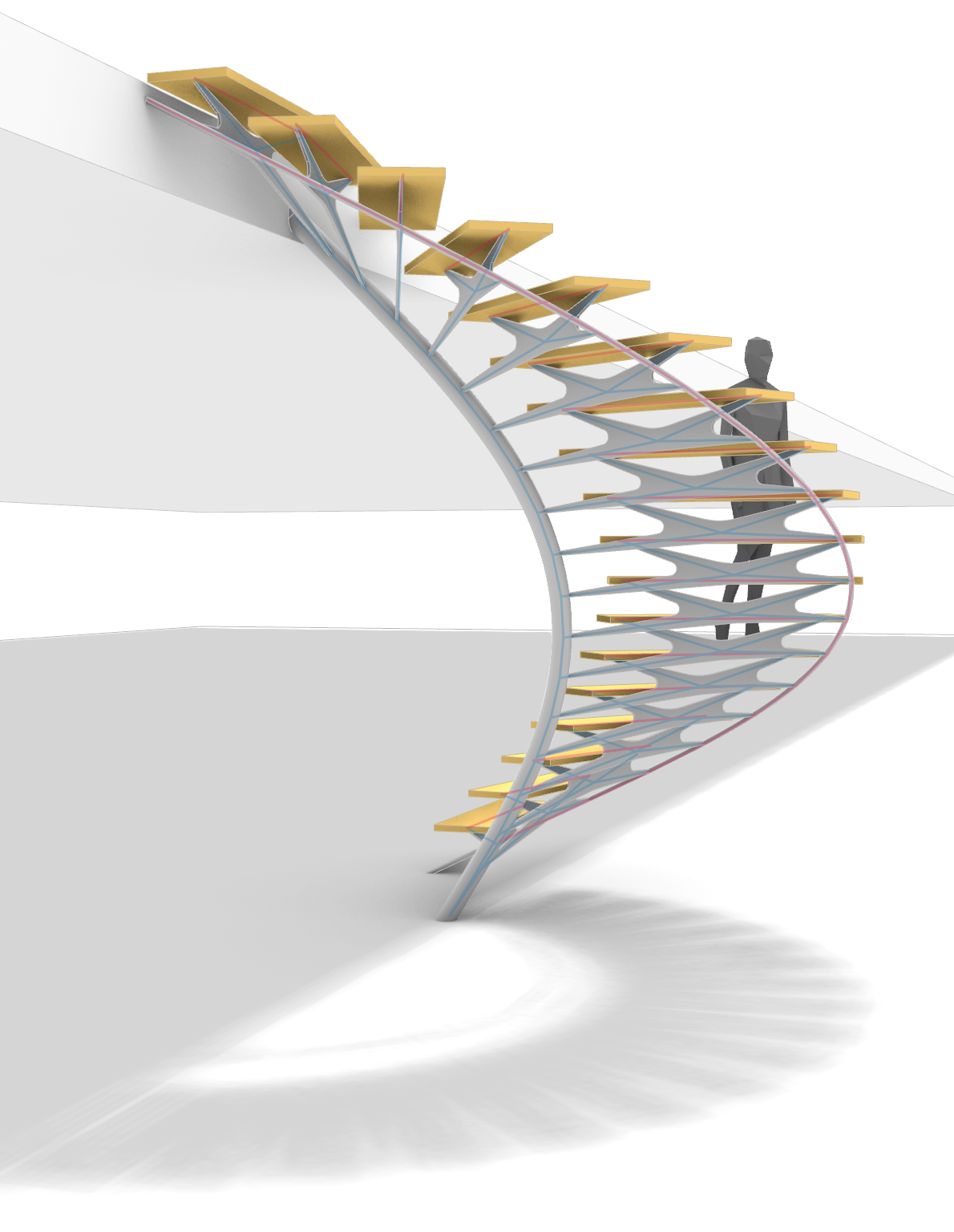}
    \caption{Architectural interpretation of the constrained form diagram $\constrainedform$ of a spiral staircase generated with the extended CEM framework.}
    \label{fig:staircase_render}
\end{figure}

Figure \ref{fig:staircase_render} finally shows one architectural interpretation of $\constrainedform$.

\section{Conclusion}
\label{conclusion}

In this paper, we presented, developed and validated three extensions to the CEM framework: auxiliary trails, AD and \texttt{compas\_cem}.
Compared to the baseline version \cite{ohlbrock_combinatorialequilibrium_2016, ohlbrock_computeraidedapproach_2020,ohlbrock_combinatorialequilibrium_2020}, our work positions the extended CEM framework as a more efficient, general and accessible approach to generate structurally efficient shapes that best meet force and geometric design constraints.

Auxiliary trails simplified the construction of valid topology diagrams for more types of structures that were difficult to be topologically modeled otherwise, such as branching structures, triangular cantilevers, and self-stressed systems.
These helper trails also made it possible to explicitly model a structure only with deviation edges while still fulfilling the topological modeling rules of the CEM form-finding algorithm.

The application of AD enabled the automatic calculation of an exact gradient value of the CEM form-finding algorithm, no longer an approximation that depended on the calibration of a step size as in previous work.
While the FD baseline and AD saw comparable performance for small constrained form-finding problems, our experiments demonstrated that calculating a constrained state of static equilibrium can be substantially faster with AD as the number of optimization parameters increases.
Using AD opens up the possibility to accelerate design exploration cycles with the extended CEM framework, especially for large constrained form-finding problems.

With \texttt{compas\_cem}, we consolidated our work into an open-source design tool.
The tool enables the formulation and the solution of constrained form-finding problems in plain and simple Python code.
Furthermore, \texttt{compas\_cem} enables designers to use the extended CEM framework on three different operating systems and on three distinct pieces of 3D modeling software.

The work presented herein has limitations.
Computing an equilibrium state where auxiliary trails are not load-bearing hinges on the solution of a constrained form-finding problem and not on a single run of the CEM form-finding algorithm.
This optimization dependency is starker when a structure is modeled entirely with deviation edges.
Consequently, the risk of using auxiliary trails is to end up with an under-parametrized or an over-constrained problem where neither equilibrium nor any other design constraint is met.
We also hypothesize that the calculation of equilibrium states for deviation-only models with the extended CEM framework may be comparable to the numerical formulation of the \emph{Update Reference Strategy} \cite{bletzinger_generalfinite_1999} and the \emph{Force Density Method} \cite{linkwitz_einigebemerkungen_1971, schek_forcedensity_1974}, and as such, form-finding deviation-only models with our approach may share their drawbacks.
A deeper investigation of this relationship is left to subsequent publications.

Future work should look into hybrid modeling strategies that guide designers to best combine auxiliary trails with standard trail and deviation edges during the construction of a topology diagram. 
Other future research directions are to add regularization terms to our current penalty approach to handle outlier constraints more robustly and to experiment with more types of objective functions as presented in \cite{cuvilliers_constrainedgeometry_2020}.
We are also interested in leveraging more complex gradient-based optimization techniques such as Newton-based optimization methods that utilize the second-order derivatives of the system solution (i.e. the Hessian) to solve constrained form-finding problems more efficiently \cite{nocedal_numericaloptimization_2006}.
By delegating the computation of derivative values to a computer using AD, we can now compose arbitrary design constraints and calculate higher-order derivatives with minimal friction: the only requisites are that the constraint and the objective functions are differentiable and written in (Python) code.

The adoption of computational techniques like AD can make gradient-based optimization more accessible to researchers in the field and it can propel the development of integrative and efficient frameworks that generate forms imbued with structural and other non-structural design requirements.
We ultimately hope our work helps to position constrained form-finding methods as viable tools to tackle practical structural design problems on a wider spectrum of structural typologies, beyond the conventional catalog of shells and cable nets.
\section*{CRediT}

\textbf{Rafael Pastrana:} Conceptualization, Methodology, Software, Validation, Investigation, Writing - Original draft, Writing - Review \& Editing, Visualization.
\textbf{Ole Ohlbrock:} Conceptualization, Methodology, Writing - Original Draft, Writing - Review \& Editing.
\textbf{Thomas Oberbichler:} Methodology, Formal analysis, Writing - Original Ddaft, Writing - Review \& Editing.
\textbf{Pierluigi D'Acunto:} Conceptualization, Writing - Review \& Editing, Supervision.
\textbf{Stefana Parascho:} Funding acquisition, Writing - Review \& Editing, Supervision.
\section*{Acknowledgements}

We thank Isabel Moreira de Oliveira from the Form-Finding Lab at Princeton University for her valuable suggestions during the development and edition of this paper.
This work was supported in part by the U.S. National Science Foundation under grant OAC-2118201 and the Deutsche Forschungsgemeinschaft, Germany (DFG) under project 434336509.


\bibliographystyle{\bibliopath/elsarticle-num}
\bibliography{\bibliopath/cem_ad, \bibliopath/manual_entries}



\end{document}